\documentclass[letterpaper,english,aps,pra,twocolumn,floatfix, showpacs, amsfonts, amssymb]{revtex4-1}
\usepackage{epsfig, graphicx,amsmath,amssymb,float}
\usepackage[T1]{fontenc}
\usepackage[latin1]{inputenc}
\usepackage{amsmath}
\usepackage{amssymb}
\usepackage{amscd}
\usepackage{bm}
\usepackage{babel}
\usepackage{color}
\usepackage{booktabs}
\usepackage{gensymb}
\usepackage{natbib}
\usepackage[T1]{fontenc}
\usepackage[colorlinks=true,linkcolor=blue,citecolor=blue,urlcolor=blue,filecolor=blue]{hyperref}
\usepackage{tabularx}

\begin{document}
\title{Fermi-Bose mixture in mixed dimensions}
\author{M. A. Caracanhas$^{1,2}$,  F. Schreck$^{3}$ and C. Morais Smith$^{1}$}
\affiliation{$^1$Institute for Theoretical Physics, Center for Extreme Matter and Emergent Phenomena, Utrecht University, Princetonplein 5, 3584 CC Utrecht, the Netherlands\\ $^2$Instituto de
F\'{i}sica de S\~ao Carlos, Universidade de S\~ao Paulo, C.P. 369, S\~ao Carlos, SP, 13560-970, Brazil \\ $^3$ Van  der  Waals-Zeeman  Institute,  Institute  of  Physics,  University  of
Amsterdam, Science  Park  904,  1098  XH  Amsterdam,  The  Netherlands}
\date{\today}

\begin{abstract} One of the challenging goals in the studies of many-body physics with ultracold atoms is the creation of a topological $p_{x} + ip_{y}$ superfluid for identical fermions in two dimensions (2D). The expectations of reaching the critical temperature $T_c$ through $p$-wave Feshbach resonance in spin-polarized fermionic gases have soon faded away because on approaching the resonance, the system becomes unstable due to inelastic-collision processes. Here, we consider an alternative scenario in which a single-component degenerate gas of fermions in 2D is paired via phonon-mediated interactions provided by a 3D BEC background. Within the weak-coupling regime, we calculate the critical temperature $T_c$ for the fermionic pair formation using the Bethe-Salpeter formalism, and show that it is
significantly boosted by higher-order diagrammatic terms, such as phonon dressing and vertex corrections. We describe in detail an experimental scheme to implement our proposal, and show that the long-sought $p$-wave superfluid is at reach with state-of-the-art experiments.
\end{abstract}


\maketitle

\section{Introduction}
The quest for the experimental realization of a chiral $p_x+ip_y$ superconductor in two dimensions (2D) is gathering increasing attention because this phase exhibits Majorana modes, which are relevant for
constructing fault-tolerant topological quantum computers \cite{kitaev,zoller}. Although a chiral $p$-wave superfluid has been shown to occur in the A-phase of $^3$He at high pressure \cite{volovik}
and experiments have revealed that Strontium ruthenate (Sr$_2$RuO$_4$) is a $p$-wave superconductor \cite{kallin},  the manipulation of the Majorana modes in these systems remains difficult. Therefore, the
prospect to create a $p$-wave superfluid using ultracold atoms is very appealing because these systems allow for great control of the degrees of freedom.

Several possibilities to generate chiral superfluids have been proposed in the context of ultracold atoms in optical lattices: by using orbital degrees of freedom \cite{sengstock,hemmerich}, spin-orbit
coupling \cite{spielman,sarma} or dipolar interaction \cite{lewenstein,shlyapnikov}. However, these methods either bring new problems to the experimental implementation, such as heating and ultracold
chemical-reactions, or require a sophisticated optical-lattice setup and further manipulations to populate the $p$-orbitals.

Here, we adopt a completely different, but feasible route to produce $p$-wave superfluids, which consists of inducing the {pairing} among the 2D polarized fermionic atoms through a 3D bath of bosonic
excitations. The dimensional mismatch between the fermions and the excitations that mediate their interaction leads to a huge increase of the superconducting gap, and consequently of the critical
temperature for the observation of the chiral superfluid. The main advantage of our proposal is that it avoids three-body losses and dynamical instabilities (phase separation), which constitute major problems in a strongly-interacting Fermi-Bose mixture.

Mixed-dimension mixtures of two-species fermions with weak interaction were investigated previously \cite{nishida,Nishida2010poa}, with the coupling between polarized fermions in 2D mediated by the particle-hole excitations of a 3D Fermi-sea background. In spite of the high stability of the Fermi-Fermi mixture, the Fermi-Bose mixture, with phonon excitations, provides much higher magnitude for the $p$-wave coupling between fermions.
Recently, a 2D-3D mixture of fermions and bosons was considered, and the Berezinskii-Kosterlitz-Thouless (BKT) critical temperature was determined accounting for effects of retardation \cite{bruun}. However, many-body effects were neglected. We argue here that the proximity between the Fermi and sound velocities requires the inclusion of many-body corrections, namely the vertex ladder-diagrams and the RPA dressing of the phonon propagator {\cite{schrieffer,Roy2014mta}.

We calculate these higher-order contributions, which are usually disregarded in the BCS treatment of conventional
superconductors, and show that they significantly contribute to increase the magnitude of the anomalous $p$-wave gap in the Fermi-Bose mixture in mixed dimensions. In this calculation, however, we do not consider the renormalization of the pole of
the Green's function, nor take into account retardation effects (the influence of the frequency of the irreducible vertex). The fermions self-energy due to the scattering of the background excitations can be neglected due to the small value of the Fermi-Bose coupling $g_{FB}$, and retardation effects should not provide a relevant contribution to the vertex \cite{kagan} because the singularity for pair formation must come from scattering in the Fermi-surface (Cooper instability \cite{schrieffer,abrikosov}).
The simultaneous analysis of both these effects, i.e., retardation and high-order vertex correction, is a tremendous task. Since our calculations are performed in the small momentum limit,  if we would consider retardation, it should enhance the positive region of the vertex because correlation between the fermions leads to an even higher prediction to the critical temperature for $p$-wave superfluid formation ($T_c^p$) \cite{Grimaldi}. Hence, the very high value of $T_c^p$ that we found due to the vertex correction is actually a lower bound, given the approximations performed.

This paper is structured as follows: Sec.~II presents the system Hamiltonian for bosonic and fermionic species, whereas in Sec.~III the interaction between  the fermions, mediated by the bosonic
excitations, is characterized. In sections~IV and V, we build the BCS Hamiltonian for the 2D system and solve the associated gap equation, respectively. Higher-order corrections for the gap magnitude are evaluated in
Sec.~VI, and the experimental feasibility, conclusions and implications of this work are discussed respectively in Sec.~VII and Sec.~VIII.

\section{System Hamiltonian}
We start by defining the Hamiltonian $\hat{H} = \hat{H}_B+\hat{H}_F+\hat{H}_{FB}$, where the boson-field operators $\hat{\phi}$ live in 3D, whereas the polarized fermions ${\hat{\psi}}$
live in 2D, (assuming $\hbar =1$)
\begin{eqnarray} \nonumber
&& \hat{H}_B=\int dz \int d^2x \hat{\phi}^{\dag}(t,\mathbf{x},z)\Big[-\frac{\nabla^2}{2m_B}  \\  && \qquad   \qquad   \quad + \frac{g_{B}}{2}
\hat{\phi}^{\dag}(t,\mathbf{x},z)\hat{\phi}(t,\mathbf{x},z)-\mu_B\Big]\hat{\phi}(t,\mathbf{x},z),  \\  \label{system}
&& \hat{H}_F=\int d^2x \hspace{0.15cm}  \hat{{\psi}}^{\dag}(t,\mathbf{x})\Big[-\frac{\nabla^2}{2m_F}-\mu_F\Big]\hat{\psi}(t,\mathbf{x}), \\  \nonumber
&&\hat{H}_{FB} \hspace{-0.1cm} =\hspace{-0.1cm}g_{FB}\hspace{-0.15cm} \int \hspace{-0.1cm} dz \hspace{-0.15cm}\int\hspace{-0.1cm}  d^2x  \delta(z)
\hat{\psi}^{\dag}(t,\mathbf{x})\hat{\phi}^{\dag}(t,\mathbf{x},z)\hat{\phi}(t,\mathbf{x},z)\hat{\psi}(t,\mathbf{x}), \\
\end{eqnarray} with the mass of the bosonic and fermionic species given by $m_B$ and $m_F$, and their chemical potentials by $\mu_B$ and $\mu_F$,  respectively. The intra- and interspecies
contact repulsive interactions are characterized by  the coupling constants $g_{B}$ and $g_{FB}$, respectively. We can  express the boson-field operators in terms of a discrete set of bosonic modes $\hat{b}_{\mathbf{q}}$, with $V$
the volume of the 3D space, \begin{equation}
\hat{\phi}(t,\mathbf{x},z) = \frac{1}{\sqrt{V}}\sum_{\mathbf{q}}e^{i \mathbf{q}\cdot\mathbf{r}} \hat{b}_{\mathbf{q}}(t),
\end{equation}
which allows us to rewrite the bosonic part of the Hamiltonian in momentum space, \begin{eqnarray} \nonumber
&&\hat{H}_B(t) = \sum_{\mathbf{q}}\left(\frac{q^2}{2m_B}-\mu_B\right)  \hat{b}^{\dagger}_{\mathbf{q}}(t) \hat{b}_{\mathbf{q}}(t)   \\  && +
\frac{g_{B}}{2V}\sum_{\mathbf{q},\mathbf{q'},\mathbf{q''}}\hat{b}^{\dagger}_{\mathbf{q}+\mathbf{q''}}(t) \hat{b}^{\dagger}_{\mathbf{q'}-\mathbf{q''}}(t) \hat{b}_{\mathbf{q}}(t)
\hat{b}_{\mathbf{q'}}(t).
\end{eqnarray}

To characterize the Bose-Einstein condensate, we now use Bogoliubov theory to deal with the macroscopic occupation of the zero-momentum state, that is $\hat{b}_0 = \hat{b}_0^{\dagger} = \sqrt{N_0}$.
Neglecting higher-order fluctuations, we obtain \begin{eqnarray} \nonumber
&&\hat{H}_B(t)  =  \frac{g_{B}N_0^2}{2V}+ \sum_{\mathbf{q}}\Big(\frac{q^2}{2m_B} + n_Bg_{B}\Big) \hat{b}^{\dagger}_{\mathbf{q}}(t) \hat{b}_{\mathbf{q}}(t)   \\  && + \frac{g_{B}n_B}{2}\sum_{\mathbf{q}}
\left[\hat{b}^{\dagger}_{\mathbf{q}}(t) \hat{b}^{\dagger}_{-\mathbf{q}}(t)+\hat{b}_{\mathbf{q}}(t) \hat{b}_{-\mathbf{q}}(t)\right].
\end{eqnarray} After symmetrizing the above expression, with a sum covering half of the momentum space, and performing a  Bogoliubov canonical transformation $\hat{b}_{\mathbf{q}}=
u_q\hat{\beta}_{\mathbf{q}}-v_q\hat{\beta}^{\dagger}_{-\mathbf{q}}$ and $\hat{b}_{-\mathbf{q}}= u_q\hat{\beta}_{-\mathbf{q}}-v_q\hat{\beta}^{\dagger}_{\mathbf{q}}$, where we select the real
parameters $u_q, v_q$ in order to have diagonal-base operators ($\hat{\beta},\hat{\beta}^{\dagger}$) for $H_B$, we find \begin{equation}
\hat{H}_B(t)  = \frac{g_{B}n_BN_0}{2}+ \sum_{\mathbf{q} (\mathbf{q}\neq0)} \omega_q \hat{\beta}^{\dagger}_{\mathbf{q}}(t) \hat{\beta}_{\mathbf{q}}(t) -\frac{1}{2}  \sum_{\mathbf{q}(\mathbf{q}\neq0)}
(\xi_q-\omega_q),
\end{equation} with the energy spectrum for the free Bogoliubov-modes excitation $\omega_q=\sqrt{\xi_q^2-(g_{B}n_B)^2}$, where \begin{equation}\xi_q=\frac{q^2}{2m_B}+g_{B}n_B.\end{equation}

Applying the same set of transformations  for the interspecies-interaction Hamiltonian ($H_{FB}$), and considering $u_q = \sqrt{{\xi_q}/{\omega_q}+1}/{\sqrt{2}}$ and
$v_q=\sqrt{{\xi_q}/{\omega_q}-1}/{\sqrt{2}}$, with $\hat{\psi}(t,\mathbf{x}) = ({1}/{\sqrt{S}})\sum_{\mathbf{p}} e^{i\mathbf{p}\cdot\mathbf{x}}\hat{a}_{\mathbf{p}}(t)$, where $S$ denotes the 2D
surface, we get \begin{eqnarray} \nonumber \label{eq9}
&&\hat{H}_{FB}(t)  =g_{FB}n_BN_F  \\ \nonumber  && +\frac{g_{FB}\sqrt{N_0}}{V}\sum{}^{'}_{\mathbf{p},\mathbf{q_{\bot}},q_z} V_q
\hat{a}^{\dag}_{\mathbf{p}}(t)\left[\hat{\beta}_{\mathbf{q}}(t)+\hat{\beta}^{\dagger}_{-\mathbf{q}}(t)\right]\hat{a}_{\mathbf{p}-\mathbf{q_{\bot}}}(t), \\
\end{eqnarray} with \vspace{-0.25cm}\begin{equation} \label{eqC}
V_q=\left(\frac{{q^2}}{{q^2}+4m_Bg_{B}n_B}\right)^{1/4}.
\end{equation}  In Eq.~(\ref{eq9}), the prime symbol in the sum indicates that $\mathbf{q}\neq0$, and we separate the components of $\mathbf{q}=(\mathbf{q}_{\bot},q_z)$, to account for momentum
conservation in the plane.

\section{Effective Interaction}
As expressed in Eq.~(\ref{system}), there is no direct interaction between the polarized fermions in $H_F$, due to the Pauli exclusion principle. We show here, however, how an indirect interaction
between fermions arises from $H_{FB}$. For that, we define the effective coupling constant $\lambda_{\textrm{eff}}$ from the four-point function $\Gamma =
\Gamma(\mathbf{p},\mathbf{p'},\mathbf{k},\mathbf{k'}; \varepsilon,\varepsilon',\nu,\nu')$ as follows  \begin{eqnarray}\nonumber \label{eq1}
\Gamma  && = \hspace{-0.6cm}\prod\limits_{\substack{i=1..4 \\ \varepsilon_i=\varepsilon,\varepsilon',\nu,\nu'}} \hspace{-0.4cm} \int dt_i e^{i\varepsilon_i t_i}\Big\langle
\hat{a}^{\dag}_{\mathbf{p}}(t_1) \hat{a}^{\dag}_{\mathbf{k}}(t_2)\hat{a}_{{\mathbf{p'}}}(t_3)\hat{a}_{{\mathbf{k'}}}(t_4) e^{-i \int d{t}   \hat{H}_{FB}({t})}\Big\rangle  \\ \nonumber && = \frac{1}{S}
i \lambda_{\textrm{eff}} \delta_{\mathbf{p}+\mathbf{k},\mathbf{p'}+\mathbf{k'}}\delta(\varepsilon+\nu-\varepsilon'-\nu')  \\  &&  \times  G_0(\mathbf{p},\varepsilon) G_0(\mathbf{p'},\varepsilon-\omega)
G_0(\mathbf{k},\nu) G_0(\mathbf{k'},\nu+\omega),
\end{eqnarray} with $G_0$ corresponding to the free-fermion propagator and $\omega = \varepsilon - \varepsilon' = \nu'-\nu$.
\begin{figure}[!ht]
  \centering
  \includegraphics[width=0.25\textwidth]{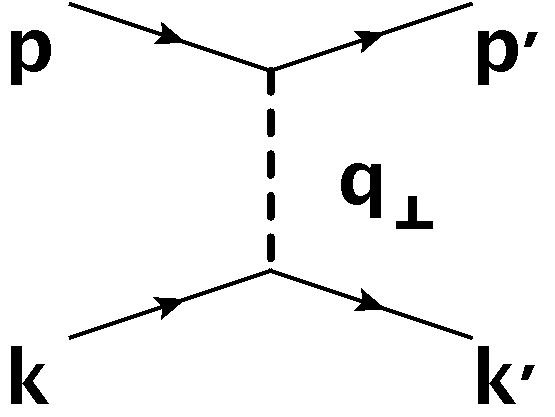}
  \caption{Second-order Feynman diagram for the interaction between two fermions in 2D induced by the Bogoliubov modes of the 3D BEC.}
  \label{fig1}
 \end{figure}

Considering the weak-coupling regime, to second order in the interaction (see Fig.~\ref{fig1}), we obtain \begin{eqnarray} \label{eq2} \nonumber
\Gamma^{(2)} && = i \frac{g^2_{FB}n_B}{V}\delta_{\mathbf{p}+\mathbf{k},\mathbf{p'}+\mathbf{k'}}\delta(\varepsilon+\nu-\varepsilon'-\nu')   \sum_{q_z} V_q^2 D_0(\mathbf{q},\omega)
 \\ && \times G_0(\mathbf{p},\varepsilon) G_0(\mathbf{p'},\varepsilon-\omega) G_0(\mathbf{k},\nu) G_0(\mathbf{k'},\nu+\omega),    \end{eqnarray} where $D_0(\mathbf{q},\omega)$ denotes the free-phonon
 propagator and $\mathbf{q}_{\bot}=\mathbf{p}-\mathbf{p'}=\mathbf{k'}-\mathbf{k}$.  Comparing Eq.~(\ref{eq1}) and Eq.~(\ref{eq2}), we find \begin{eqnarray}\label{eq3} \nonumber
\lambda_{\textrm{eff}} =  g_{FB}^2 n_B \int_{-\infty}^{\infty} \frac{dq_z}{2\pi}  \left(\frac{\frac{q^2}{2m_B}}{\frac{q^2}{2m_B}+2g_{B}n_B}\right)^{1/2} \hspace{-0.25cm}
\frac{2\omega_q}{\omega^2-\omega_q^2+i\delta} . \\
\end{eqnarray}

For low-energy processes, where the scattered fermions are kept around the 2D Fermi surface, we can assume $\omega \sim 0$, and Eq.~(\ref{eq3}) can be simplified as \begin{eqnarray}\nonumber
\label{eq0}
\lambda_{\textrm{eff}} \; &&= - \frac{2}{\pi} m_B g_{FB}^2 n_B \int_{-\infty}^{\infty} dq_z  \frac{1}{q_z^2+{q_{\bot}}^2+4m_Bg_{B}n_B} \\
&&= - 2 m_B g_{FB}^2 n_B   \frac{1}{\sqrt{{q_{\bot}}^2+4m_Bg_{B}n_B}}.
\end{eqnarray} Hence, an effective potential $\lambda_{\textrm{eff}} = V_{\textrm{eff}}(q_{\bot}= |\mathbf{p'}-\mathbf{p}|)$ is generated between the fermions, as a function of the momentum exchange
$\mathbf{Q}$ between the scattered particles. In 2D real space, with coordinate $\mathbf{R}$, this yields an attractive Yukawa potential between the fermionic particles in the plane,
\begin{eqnarray}\nonumber
V_{\textrm{eff}}(\mathrm{R}) =  \int d^2Q e^{i  \mathbf{Q}\cdot\mathbf{R}} V_{\textrm{eff}} (Q) \\
= - 2\pi \frac{g_{FB}^2}{g_{B}} \frac{1}{\xi^2}    \frac{1}{\mathrm{R}} e^{- \frac{\sqrt{2}}{\xi} \mathrm{R}},
\end{eqnarray} with range given by the healing length $\xi = 1/\sqrt{2m_B g_{B} n_B}$ of the BEC.

\section{BCS Hamiltonian}
We consider the generalized BCS-type Hamiltonian in momentum space for the fermions in the plane, \begin{eqnarray} \nonumber \label{eq4}
&&\hat{H}^{\prime}_F = \hspace{-0.2cm}\int\hspace{-0.2cm}  \frac{d^2p}{(2\pi)^2}\hspace{-0.05cm} \bigg\{\hspace{-0.1cm} \left(\hspace{-0.05cm} \frac{p^2}{2m_F}-\mu\hspace{-0.05cm}\right)\hspace{-0.1cm}
\hat{a}^{\dag}(\mathbf{p})\hat{a}(\mathbf{p}) +\frac{1}{2}\hspace{-0.1cm} \int \hspace{-0.1cm} \frac{d^2k d^2k'}{(2\pi)^4} V_{\textrm{eff}}(\mathbf{p},\mathbf{k}) \\ &&\times
\hat{a}^{\dag}\hspace{-0.05cm} \left({\mathbf{k'}}/{2}+\mathbf{k}\right)\hspace{-0.05cm}\hat{a}^{\dag}\hspace{-0.05cm} \left({\mathbf{k'}}/{2}-\mathbf{k}\right)\hspace{-0.05cm} \hat{a}\hspace{-0.05cm}
\left({\mathbf{k'}}/{2}-\mathbf{p}\right)\hspace{-0.05cm} \hat{a}\hspace{-0.05cm} \left({\mathbf{k'}}/{2}+\mathbf{p}\right) \bigg\}, \end{eqnarray}  with a momentum-dependent  mediated interaction
$V_{\textrm{eff}}(\mathbf{p},\mathbf{k})$ and $\mu = \mu_F-n_Bg_{FB}$. According to Eq.~(\ref{eq0}), we consider the interaction potential  \begin{eqnarray} \label{veff0}
V_{\textrm{eff}}(\mathbf{p},\mathbf{k}) = - V_0\frac{1}{\sqrt{|\mathbf{p}-\mathbf{k}|^2+2\xi^{-2}}}, \end{eqnarray} with $V_0 = 2 g_{FB}^2n_Bm_B$. After symmetrizing the BCS Hamiltonian properly, we apply the Bogoliubov transformation and find a new basis of operators (see App.~\ref{apA} for details) to build the diagonal form \begin{eqnarray} && \hat{H}_F^{BCS} \nonumber
 = {\sum_{\mathbf{p}}}   E_{p} \hat{\alpha}^{\dag}_{\mathbf{p}}\hat{\alpha}_{\mathbf{p}} + \\  &&+
 \frac{1}{2}{\sum_{\mathbf{p}}}\bigg\{\frac{|{\triangle}_{\mathbf{p}}|^2}{E_{p}}\Big[1-2n_F(E_{p})\Big]+ \left(\epsilon_{p}-E_{p}\right)\bigg\},\end{eqnarray} with the energy dispersion $E_{p}=\sqrt{\epsilon_{p}^2+|{\triangle}_{\mathbf{p}}|^2}$ and the occupation function $n_{F}(E_{p}) = [\exp(\beta E_{p}) +1]^{-1}$ of the Bogoliubov modes, where $\beta=(k_BT)^{-1}$. As shown in App.~\ref{apA}, now we can also write the gap in terms of the mean value over this new basis, to obtain
\begin{eqnarray}\label{eq5}
\triangle_{\mathbf{p}} = - \int \frac{d^2k}{(2\pi)^2} V_{\textrm{eff}}(\mathbf{p},\mathbf{k}) \frac{{\triangle}_{\mathbf{k}}}{2E_{k}}\Big[1-2n_F(E_{k})\Big].\quad
\end{eqnarray}

\section{GAP Equation}
To solve the integral equation for a momentum-dependent pairing gap in Eq.~(\ref{eq5}), it is convenient to use the 2D partial-wave expansion of the effective potential \cite{anderson,chubukov},
\begin{eqnarray} \label{eq6}
V_{\textrm{eff}}(\mathbf{p},\mathbf{k}) = \sum_{\ell}  V_{\textrm{eff}}^{(\ell)}(p,k) \cos[\ell (\theta -\varphi)],
\end{eqnarray} with $\ell$ integer, $p = |\mathbf{p}|$,  $k = |\mathbf{k}|$, and where we associated the angles ${\theta}_{\mathbf{\hat{p}}}=\theta$ and ${\theta}_{\mathbf{\hat{k}}}=\varphi$. Because we are assuming low-energy processes, with the
scattered momentum close to the Fermi surface, it is reasonable to consider $p\sim k = k_{F}$ in the coefficients of Eq.~(\ref{eq6}). For $\ell=1$, considering the even parity of the potential, we have
\begin{eqnarray}\nonumber \label{l1}
V_{\textrm{eff}}^{(1)}(k_{F})  &&= \frac{1}{\pi^2}\int \hspace{-0.2cm}\int_{-\pi}^{\pi} \frac{-V_0  \cos\varphi  \cos\theta }{\sqrt{2\xi^{-2}+2k_F^2\left[1-\cos(\theta-\varphi)\right]}} d\theta
d\varphi \\
&&= \frac{2\sqrt{2}}{\pi}\; V_0 \xi \; \mathcal{F}(k_F\xi),
 \end{eqnarray} where \begin{eqnarray} \label{eqF}
&&\mathcal{F}(X)= \frac{E[-2 X^2] - (1 + X^2) \; K[-2 X^2]}{X^2},
\end{eqnarray} \\ with $E[X]$ the complete elliptic integral, $K[X]$ the complete elliptic integral of the first kind, and $X= k_F\xi$ (see the inset of Fig.~\ref{fig2}). \begin{figure}[!ht]
  \centering
  \includegraphics[width=0.45\textwidth]{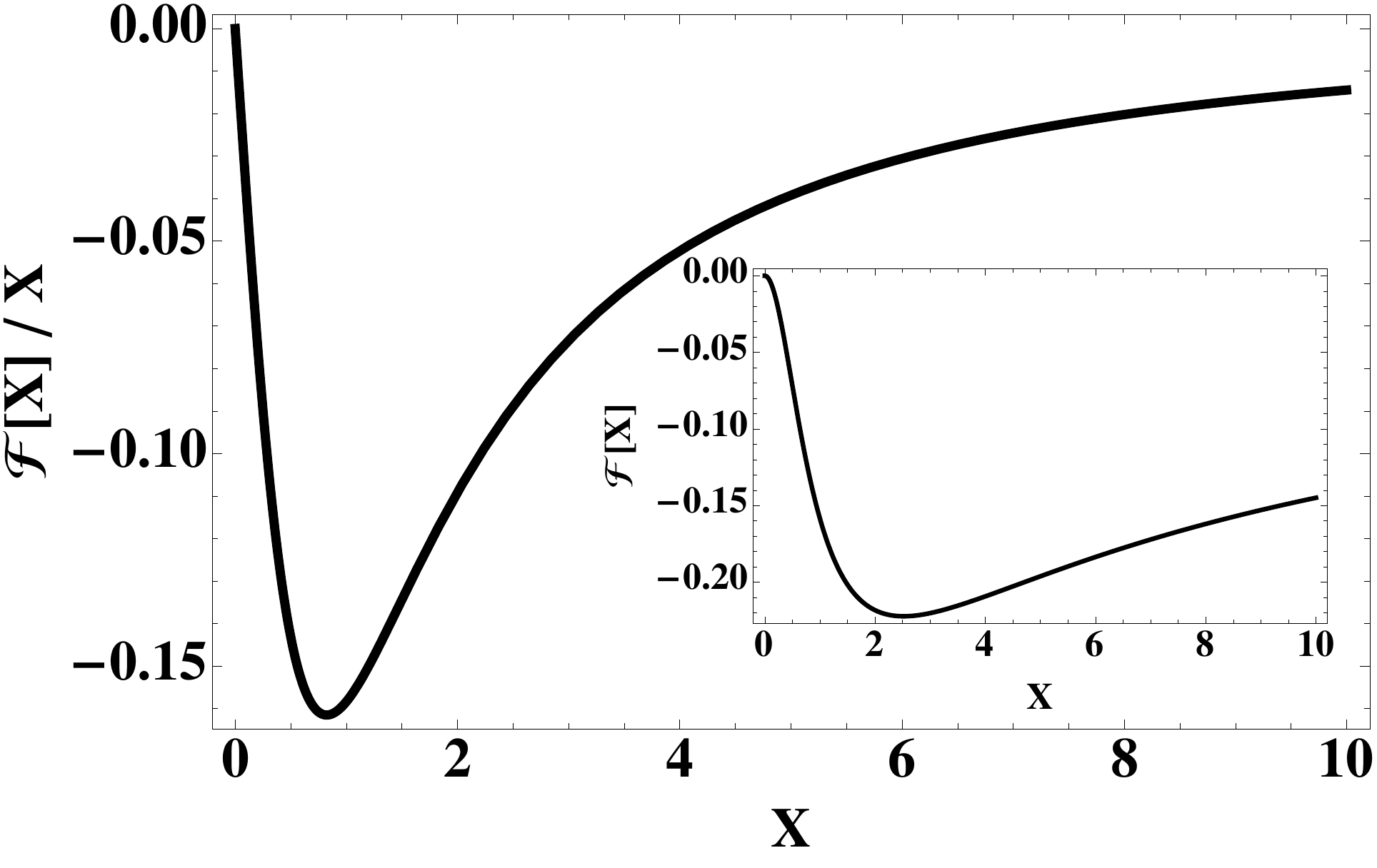}
  \caption{Profile of the function $\mathcal{F}(X)/X$ used to estimate the maximum gap in Eq.~(\ref{GAPmax}). Inset: harmonic $\ell=1$ of the effective potential, i.e. $\mathcal{F}(X)$ in Eq.~(\ref{l1}),  as a function of $X=k_F\xi$.}
   \label{fig2}
\end{figure}  Since in the weak-coupling limit one expects that the mixing of different angular momentum $\ell$ will be small, we are in a position to solve the gap equation by applying the pure
$\ell$-type ansatz  $\triangle_{\mathbf{p}} = \triangle^{(\ell)} e^{i\ell\theta_{\hat{\mathbf{p}}}}$  in Eq.~(\ref{eq5}). That gives \begin{eqnarray} \nonumber \label{eq7} \nonumber
 \triangle^{(\ell)} e^{i\ell\theta_{\hat{\mathbf{p}}}} &=& - \int \frac{d^2k}{(2\pi)^2} V_{\textrm{eff}}(\mathbf{p},\mathbf{k})\frac{\triangle^{(\ell)}
 e^{i\ell\theta_{\hat{\mathbf{k}}}}}{2E_{k}}\left[1-2n_F(E_{k})\right]
\\ \nonumber   1 &=& - \int \frac{k dk d\varphi}{(2\pi)^2} \sum_{\ell^{\prime}}  V_{\textrm{eff}}^{(\ell^{\prime})}(k_F) \cos[\ell^{\prime} (\theta-\varphi)] \\ &\times&
\frac{e^{i\ell(\varphi-\theta)}}{2E_{k}}\left[1-2n_F(E_{k})\right].
\end{eqnarray} Analytical solutions for $\triangle^{\textrm{Max}}$ and $T_c$ can be obtained in two limiting cases: 1)$\;T \rightarrow 0$, where we have the maximum gap value, and 2) $\;T \rightarrow T_c$, where
the gap goes to zero. For the first limit, we find $E_{k} = \sqrt{\epsilon_{k}^2+|\triangle^{(\ell)}|^2}$ and $n_F(E_{k}) \rightarrow 0$. Then, applying the orthogonality condition given by the angular
integral of equation (\ref{eq7}), we eliminate the sum in $\ell'$ to obtain \begin{eqnarray} \nonumber \label{eq8}
&& 1 = - \frac{1}{(2\pi)^2} \frac{\pi}{4}V_{\textrm{eff}}^{(\ell)}(k_F) \int k dk  \frac{1}{\sqrt{\epsilon_{k}^2+|\triangle^{(\ell)}|^2}}
\\ && 1 = - \frac{1}{2\pi} V_{\textrm{eff}}^{(\ell)}(k_F) \frac{\pi}{4} \frac{m_F}{2\pi}\int_{0}^{\Lambda_{\varepsilon}} d\varepsilon   \frac{1}{\sqrt{\varepsilon^2+|\triangle^{(\ell)}|^2}},
\end{eqnarray} where we can identify the density of states in the Fermi surface $\rho_{2D} = m_F/2\pi$ and the cut-off energy scale given by the Fermi energy of the 2D system $\Lambda_{\varepsilon}\sim
k_F^2/2m_F$. Since we consider the small-momentum regime, the fermions are scattered to states around the Fermi level. As can be seen from Table~\ref{TabLiYbParameters} in the  experimental section, $k_F$ is very close to the healing length ($\xi^{-1}$), which characterizes the range of the interaction potential.

One can show that the induced attraction Eq.~(\ref{veff0}) is strongest in the $p$-wave channel. That means that the dominant pairing instability is in the channel with orbital angular
momentum $\ell=1$, and the most stable low-temperature phase, or with highest critical temperature, has $p_{x} + ip_{y}$ symmetry \cite{anderson, nishida}. We can then solve Eq.~(\ref{eq8}) for the
maximum gap \begin{eqnarray} \label{GAP}
\triangle^{\textrm{Max}} = \triangle^{(1)}= 2 \Lambda_{\varepsilon} \; \exp\bigg(\frac{1}{\rho_{2D}\tilde{V}_{\textrm{eff}}^{(1)}(k_F)}\bigg),
\end{eqnarray} \\ with $\;\tilde{V}_{\textrm{eff}}^{(1)}(k_F) = {V}_{\textrm{eff}}^{(1)}(k_F)/8$.

The vertex renormalization for two particles in vacuum allows us to express the bare coupling parameter as $g_{FB} \rightarrow -2\pi a_{\textrm{eff}} /\sqrt{m_Bm_{FB}}$  \cite{nishida2}, with the reduced
mass $m_{FB}=m_Bm_F/(m_B+m_F)$ and the effective two-body scattering length $a_{\textrm{eff}}$ for a 2D-3D scattering. The latter will be a function of the original 3D scattering length $a_{FB}$ and of
the axial confinement. That gives \begin{eqnarray}
 &&\tilde{V}_{\textrm{eff}}^{(1)}(k_F) = 2\sqrt{2}\pi \frac{n_B a_{\textrm{eff}}^2\xi}{m_{FB}} \mathcal{F}(k_F\xi).
\end{eqnarray} Considering $k_F=\sqrt{4\pi n_F}$ and $\xi = 1/\sqrt{8\pi n_Ba_B}$, we get the variable \begin{eqnarray}
\xi k_F=  \frac{1}{\sqrt{2}} \sqrt{\frac{n_F}{a_Bn_B}}.
\end{eqnarray} Thus, we estimate the gap in Eq.~(\ref{GAP}) using \begin{eqnarray} \label{maxgap}\rho_{2D} \tilde{V}_{\textrm{eff}}^{(1)}(k_F) = \frac{\sqrt{2}}{8\pi}
\frac{m_F}{m_{FB}}\frac{a_{\textrm{eff}}^2k_F}{a_B}\frac{\mathcal{F}(k_F\xi)}{k_F\xi}.
\end{eqnarray} For $a_B n_B^{1/3} \sim 0.01$ and $a_{\textrm{eff}}k_F\sim 0.1$, we consider the maximum value for $\rho_{2D} |\tilde{V}_{\textrm{eff}}^{(1)}(k_F)|$ with $\mathcal{F}(X)/X \sim -0.15$, restricting $X$ in the interval $[0.5-1.5]$ (see Fig.~\ref{fig2}), to determine \cite{calculo} \begin{eqnarray} \label{GAPmax}
\triangle^{\textrm{Max}} \sim 0.01 \Lambda_{\varepsilon}.
\end{eqnarray}

\section{Higher order correction to the Effective 2D-3D interaction}
The previous section shows how to optimize the gap
value by manipulating the condensate density, which controls
the magnitude and range of the induced potential. In addition, the importance of choosing
an appropriate combination of the fermion and boson
atomic masses (lighter bosonic species) to maximize the
gap became clear. This issue will be further explored in Sec.~VII.

By choosing the Fermi wavelength and the healing length such that $\xi k_F\sim1$,  the Bogoliubov-sound ($c_s$) and the Fermi velocities ($v_F$) will also have close values. That requires the inclusion of higher-order {diagrammatic} terms in our ultracold-atoms model, which are usually disregarded in BCS studies.

In the following, we calculate the four-point function to $4th$ order in the interaction constant $g_{FB}$
\small \begin{eqnarray} \nonumber
\Gamma(\{\mathbf{k}_i,\tau_i\})&&=- \left\langle T_{\tau} \hat{a}_{\mathbf{k}_1}(\tau_1)\hat{a}_{\mathbf{k}_2}(\tau_2)\hat{a}^{\dagger}_{\mathbf{k}_3}(\tau_3)\hat{a}^{\dagger}_{\mathbf{k}_4}(\tau_4)
e^{-\int_0^{\beta}\hspace{-0.1cm}d\tau \hat{H}_{int}(\tau)}\right\rangle. \\
\end{eqnarray}  \normalsize We start with the interaction between the fermions in 2D and the ``phonons'' of the BEC in 3D  as given by Eq.~(\ref{eq9}) and Eq.~(\ref{eqC}).
Using the finite temperature formalism with the Matsubara Green's functions, the effective interaction between the fermions in 2D is given by  \small \begin{eqnarray} \nonumber
\Gamma_{\textrm{eff}}(\{\mathbf{k}_i,\nu_i\})=\lambda_{\textrm{eff}}\frac{\beta}{S}\delta_{\mathbf{k}_1+\mathbf{k}_2,\mathbf{k}_3+\mathbf{k}_4}\delta_{\nu_1+\nu_2,\nu_3+\nu_4} \hspace{-0.2cm}
\prod_{i=1...4}  \hspace{-0.2cm}\mathcal{G}_0(\mathbf{k}_i,\nu_i), \\ \end{eqnarray}  \normalsize with the free-fermion propagator $\mathcal{G}_0$. As seen before, the second-order expansion in the
coupling $g_{FB}$ provides \small \begin{eqnarray} \nonumber
&& \Gamma^{(2)}(\{\mathbf{k}_i,\nu_i\})= \frac{\beta}{V} g_{FB}^2 n_B \delta_{\mathbf{k}_1+\mathbf{k}_2,\mathbf{k}_3+\mathbf{k}_4}  \delta_{\nu_1+\nu_2,\nu_3+\nu_4}  \\ \nonumber && \times \sum_{q_z}
V^2_{\mathbf{q}} \mathcal{D}_0(\mathbf{q},\nu_1-\nu_4) \prod_{i=1...4}  \mathcal{G}_0(\mathbf{k}_i,\nu_i) \\ \nonumber
&& =  \hspace{-0.1cm}  \frac{-2 g_{FB}^2 n_B m_B}{\sqrt{|\mathbf{k}_1-\mathbf{k}_4|^2+2\xi^{-2}}} \frac{\beta}{S}
\delta_{\mathbf{k}_1+\mathbf{k}_2,\mathbf{k}_3+\mathbf{k}_4}\delta_{\nu_1+\nu_2,\nu_3+\nu_4} \hspace{-0.2cm} \prod_{i=1...4}  \hspace{-0.2cm}\mathcal{G}_0(\mathbf{k}_i,\nu_i),   \\ \end{eqnarray}
\normalsize where $\mathbf{q}\equiv(\mathbf{k}_1-\mathbf{k}_4,q_z)$
 and we applied the static limit to the Bogoliubov-mode propagator $\mathcal{D}_0$.

Within a higher-order expansion, we obtain the self-energy bubble diagram (see the details of the calculation in App.~\ref{apB}) \small \begin{eqnarray} \nonumber
&&\Gamma_{RPA}^{(4)}(\{\mathbf{k}_i,\nu_i\})= \hspace{-0.1cm}  \frac{4 g_{FB}^4 n_B^2 m_B^2}{|\mathbf{k}_1-\mathbf{k}_4|^2+2\xi^{-2}}\hspace{-0.1cm} \sum_{\mathbf{p}} \hspace{-0.1cm}
\frac{n_F(\epsilon_{\mathbf{p}})-n_F(\epsilon_{\mathbf{p}+\mathbf{k}_4-\mathbf{k}_1})}{\nu_4-\nu_1+\epsilon_{\mathbf{p}}-\epsilon_{\mathbf{p}+\mathbf{k}_4-\mathbf{k}_1}} \\ && \times  \frac{\beta}{S^2}
\delta_{\mathbf{k}_1+\mathbf{k}_2,\mathbf{k}_3+\mathbf{k}_4}\delta_{\nu_1+\nu_2,\nu_3+\nu_4} \prod_{i=1...4} \mathcal{G}_0(\mathbf{k}_i,\nu_i),  \end{eqnarray} \normalsize where we identify the static
polarization-bubble diagram in 2D \begin{eqnarray}
 &&P_0(\mathbf{k}_1,\mathbf{k}_4)  = \frac{1}{S} \sum_{\mathbf{p}}
 \frac{n_F(\epsilon_{\mathbf{p}})-n_F(\epsilon_{\mathbf{p}+\mathbf{k}_4-\mathbf{k}_1})}{\epsilon_{\mathbf{p}}-\epsilon_{\mathbf{p}+\mathbf{k}_4-\mathbf{k}_1}}.
\end{eqnarray} For $|\mathbf{k}_1-\mathbf{k}_4| < 2 k_F $, i.e., the external momenta in the Fermi surface, we can easily calculate the RPA series, which yields \begin{eqnarray}  \nonumber
\lambda_{\textrm{eff}}^{RPA} && = \lambda_0+\lambda_0^2 P_0+\lambda_0^3 P_0^2+ ... \\
&& =  \lambda_0 [ 1+ \lambda_0 P_0+\lambda_0^2 P_0^2+ ...],
\end{eqnarray} \\ where we defined $\lambda_0 =  - {V_0 }/{\sqrt{|\mathbf{k}_1-\mathbf{k}_4|^2+2\xi^{-2}}}$ and $P_0=-{m_F}/{2\pi}=-\rho_{2D}$. For $ \lambda_0 P_0<1$, we find \begin{eqnarray}
\label{RPA}\nonumber
\lambda_{\textrm{eff}}^{RPA} = \frac{\lambda_0}{1-\lambda_0P_0} =  \frac{-V_0}{\sqrt{|\mathbf{k}_1-\mathbf{k}_4|^2+2\xi^{-2}} - {V_0 \rho_{2D}}}. \\
\end{eqnarray}  Replacing Eq.~(\ref{veff0}) by the effective potential coming from the RPA correction in Eq.~(\ref{RPA}), we obtain an increase in the gap magnitude, as predicted by Eq.~(\ref{GAP}) (see also App.~\ref{apB} and Fig.~\ref{fig3}). \begin{figure}[!ht]
\centering
\includegraphics[width=0.45\textwidth]{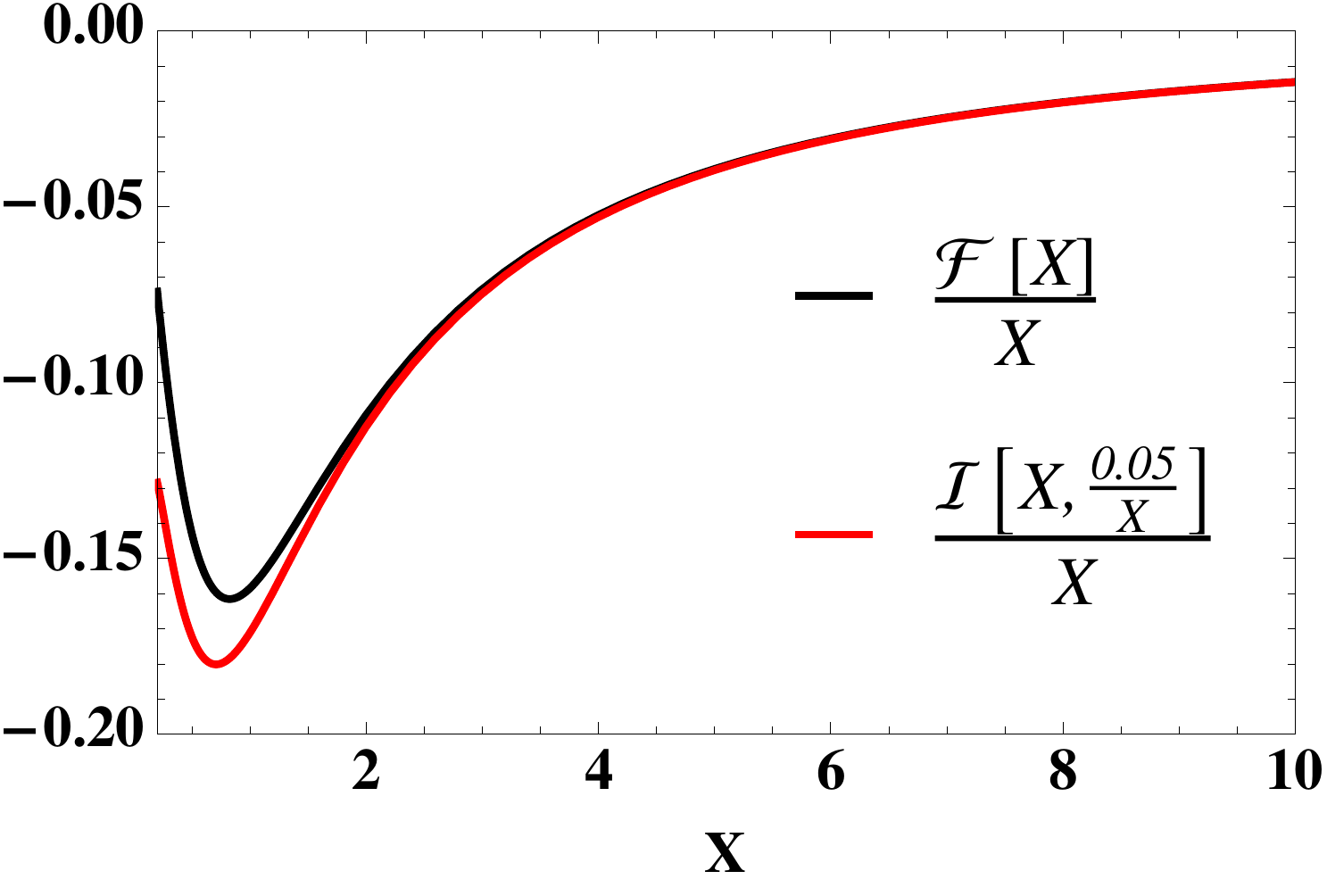}
\caption{RPA correction to the $\ell=1$ component of the effective potential, according to Eq.~(\ref{l1}) and Eq.~(\ref{eqD}).}\label{fig3}
\end{figure} Since we consider $\lambda_0 P_0$ smaller than one, we do not expect any phase instability driven by a divergence of $\lambda_{\textrm{eff}}^{RPA}$ caused by the vanishing of the denominator of Eq.~(\ref{RPA}).

The critical condition given by Eq.~(\ref{GAP}) can be obtained alternatively through the singularity in the effective interaction, which appears when the total vertex function is calculated in
the Fermi surface, considering small total momentum of the colliding particles  \cite{abrikosov,pitaevskii,chubukov}. In this case, the $\ell$-th harmonic in the exponent of Eq.~(\ref{GAP}) will be associated with the irreducible part of the vertex. Here, we determined its $\ell=1$
projection solving the Bethe-Salpeter integral equation for the ladder-series contribution. To build the series, we start with the $4th$-order vertex-correction, which reads \small
\begin{eqnarray} \nonumber \label{vertx1}
&& \Gamma_{V}^{(4)}(\{\mathbf{k}_i,\nu_i\}) =   \frac{2 g_{FB}^4 n_B^2 m_B}{\sqrt{|\mathbf{k}_1-\mathbf{k}_4|^2+2\xi^{-2}}} \frac{1}{V} \sum_{\mathbf{p},q_z} \frac{q}{\sqrt{q^2+2\xi^{-2}}} \\
\nonumber
&&\times\bigg[\frac{1}{(\omega_{\mathbf{q}}+\epsilon_{\mathbf{p}})(\omega_{\mathbf{q}}+\epsilon_{\mathbf{p}+\mathbf{k}_4-\mathbf{k}_1})}
+\frac{4n_F(\epsilon_{\mathbf{p}})\omega_{\mathbf{q}}}{(\epsilon_{\mathbf{p}}-\epsilon_{\mathbf{p}+\mathbf{k}_4-\mathbf{k}_1})(\omega^2_{\mathbf{q}}-\epsilon^2_{\mathbf{p}})}  \bigg]  \\ && \times
\frac{\beta}{S} \delta_{\mathbf{k}_1+\mathbf{k}_2,\mathbf{k}_3+\mathbf{k}_4} \delta_{\nu_1+\nu_2,\nu_3+\nu_4} \prod_{i=1...4} \mathcal{G}_0(\mathbf{k}_i,\nu_i),  \end{eqnarray} \normalsize with
$\omega_{\mathbf{q}}=\frac{q}{2m_B}\sqrt{q^2+2\xi^{-2}}$ and $\mathbf{q}\equiv(\mathbf{k}_3-\mathbf{p},q_z)$.  \normalsize The first term of Eq.~(\ref{vertx1}) is related to single-particle behavior,
i.e., the scattering of real phonons, whereas the second term corresponds to virtual phonon processes. Only the latter will be relevant in our calculation, which deals with the many-body effects with
the 2D momentum integration performed near the Fermi surface.

To evaluate the  irreducible-vertex part around the Fermi surface, perturbation theory turns out to be insufficient and we must sum the whole ladder series of diagrams, with terms proportional to the ratio $c_s/v_F$. The resulting self-consistent vertex equation is presented and solved in the App.~\ref{apB}, after performing a partial expansion of the effective interaction
$\lambda_{\textrm{eff}}^{V}$ in terms of the angular components $\lambda(|\mathbf{k}_4-\mathbf{k}_1|) = \sum_{\ell}  \lambda^{(\ell)}(k_F) \cos[\ell (\theta_4 -\theta_1)]$  \cite{chubukov,pitaevskii},
which breaks the integral equation for the total pairing vertex to a set of decoupled algebraic equations for its partial components. Finally, we obtain the vertex correction for the component
$\ell=1$   \begin{eqnarray}\label{Vc} {\lambda_{\textrm{eff}}^V}^{\hspace{-0.05cm}(1)}(k_F) =  \frac{V_{\textrm{eff}}^{(1)}(k_{F})}{1+ \frac{1}{4}V_{\textrm{eff}}^{(1)}(k_{F})\rho_{2D}
\frac{\mathcal{J}[X]}{\mathcal{F}[X]X^2 \sqrt{1+2X^2}}},
\end{eqnarray} \\ where we defined $\mathcal{J}[X] = (1+2X^2) E\left[1 - \frac{1}{1+2X^2}\right] - (1+X^2) K\left[1 - \frac{1}{1+2X^2}\right]$. Remarkably, $\frac{\mathcal{J}[X]}{\mathcal{F}[X]X^2
\sqrt{1+2X^2}} = 1!$   Including the correction given by Eq.~(\ref{Vc}) into the gap equation, according to Eq.~(\ref{GAP}),  we get \begin{eqnarray} \label{GAPF} \nonumber
\triangle_V^{\textrm{Max}}
\nonumber
&&= 2 \Lambda_{\varepsilon} \; \exp\bigg(\frac{8}{\rho_{2D}V_{\textrm{eff}}^{(1)}}+ 2\bigg)  \\
&& \sim 7.4 \; \triangle^{\textrm{Max}}.
\end{eqnarray} This is the main result of this paper: the inclusion of higher-order diagrams, usually neglected due to their complexity, actually increases the $p$-wave gap by one order of magnitude and brings it to the verge of experimental possibilities.

\section{Experimental implementation}

We now discuss the experimental feasibility of our proposal. We first examine which quantum gas mixtures are suitable to implement it, then present a scheme for a mixed-dimensional trap, and finally we summarize the experimental proposals to detect a $p$-wave superfluid.

\subsection{Mixture choice}

The most important criterion to choose the mixture is that the critical temperature for $p$-wave superfluidity $T_c^p$ has to be experimentally reachable \cite{McKay2011cis}. As guidance, we note that BECs have been evaporatively cooled to $T=0.02 T_c^{\rm BEC} = 1\,$nK \cite{Olf2015tac} and Fermi gases with $T/T_F \leq 0.05$ have been reached \cite{Navon2010teo}. We maximize $T_c^p/T_F=\gamma \triangle_V^{\rm Max}/T_F$ \cite{schrieffer} under constraints imposed by the validity of our theory and experimental constraints ($\gamma$: Euler's constant $\simeq 0.57$). The static approximation requires that $\alpha=v_F/c_s \lesssim 1$ {\cite{Migdal,Roy2014mta}}. In addition, since the effective potential has been obtained within a perturbative treatment, it is necessary that $\gamma_{eff}^2<(8\pi\gamma_{BEC})^{1/2}$. Hence, the boundaries of validity of our theoretical studies request $\gamma_{BEC}=a_B n_B^{1/3} \gtrsim 10^{-3}$ \cite{bruun} and $\gamma_{\rm eff}=a_{\rm eff} n_B^{1/3} \lesssim (8\pi\gamma_{BEC})^{1/4} \approx 0.4$. To be in the superfluid regime we finally require $T^p_c<T_{\rm KT}$, where $T_{\rm KT}$ is the Kosterlitz-Thouless transition temperature \cite{Fisher,Dalibard}.  Since $T_c^p/T_F=8.42 \exp(-1/|\rho_{2D}\tilde{V}_{\textrm{eff}}^{(1)}|)$ increases monotonically with $Y=|\rho_{2D}\tilde{V}_{\textrm{eff}}^{(1)}|$ it is sufficient to maximize $Y$, which can be expressed as \begin{eqnarray}\label{rho2DVefftilde}
Y=\frac{1}{\sqrt{4\pi}}\left(1 + \frac{m_F}{m_B}\right)\frac{\gamma_{\rm eff}^2}{\sqrt{\gamma_{BEC}}} \left|\mathcal{F}(X)\right|,
\end{eqnarray} with $X=\alpha(m_F/m_B)/\sqrt{2}$. For large $Y$, a high mass ratio $m_F/m_B$ should be selected, provided that $\alpha$ is chosen close to $\alpha_{\rm max}=3.56\,m_B/m_F$, which maximizes $|\mathcal{F}(X)|$. Since $T_F=(2\pi \hbar^2/k_B)(m_F/m^2_B)n_B^{2/3}\alpha^2\gamma_{BEC} \propto \alpha^2$, we chose in the following a slightly higher value, $\alpha = 1.5\, \alpha_{\rm max}$, which barely decreases $|\mathcal{F}(X)|$, but more than doubles $T_F$. Furthermore, a low value of $\gamma_{BEC}$ is desired and we chose a value close to its minimum. Finally, a high value of $\gamma_{\textrm{eff}}$ has to be achieved. In order to increase $\gamma_{\rm eff}$, we opt for the rather high value of $n_B=6\times 10^{14}\,$atoms/cm$^{-3}$ and the relatively low value of $a_{\rm eff}=204\,a_0$, where $a_0$ is the Bohr radius. The motivation for choosing a large density is that $T_F$ increases with $n_B$. On the other hand, low values of $a_{\rm eff}$ are more likely available in experiments than large values, and they can be reached without Feshbach or confinement induced resonances. Far from the resonances, the scattering length is given approximately by $a_{\rm eff} \sim \sqrt{m_B/m_{FB}}\,a_{\rm FB}$ \cite{Massignan2006tds,nishida2,Lamporesi2010sim}.

Further limitations arise from experimental constraints. In our scheme, a few thousand fermions will be sympathetically cooled by a much larger bath of evaporatively cooled bosons. To effectively implement evaporative and sympathetic cooling, a sufficient rate of elastic collisions and low rates of heating and loss are required. These conditions limit the range of suitable interaction properties, the gas densities, and the trap designs. An upper limit on $n_B$ is imposed by the requirement to keep the BEC in the 3D regime for the finite number of bosons available. A lower limit on $a_B$ is imposed by the requirement of a sufficient elastic collision rate between bosons $\Gamma_{\rm el,B} \propto n_B a_B^2$. Together, these requirements lead to an additional, experimental, lower limit on $\gamma_{BEC}$. Attention has also to be given to the rate of 3-body losses involving one fermion and two bosons ($\Gamma_{\rm FBB}\propto n_B^2 a_{\rm FB}^4$ \cite{Esry2008,Laurent2017}), even considering the important role played by the mixed dimensionality in inhibiting the interspecies molecular formation \cite{nishida}.
\begin{figure*}[!ht]
  \centering
  \includegraphics[width=\textwidth]{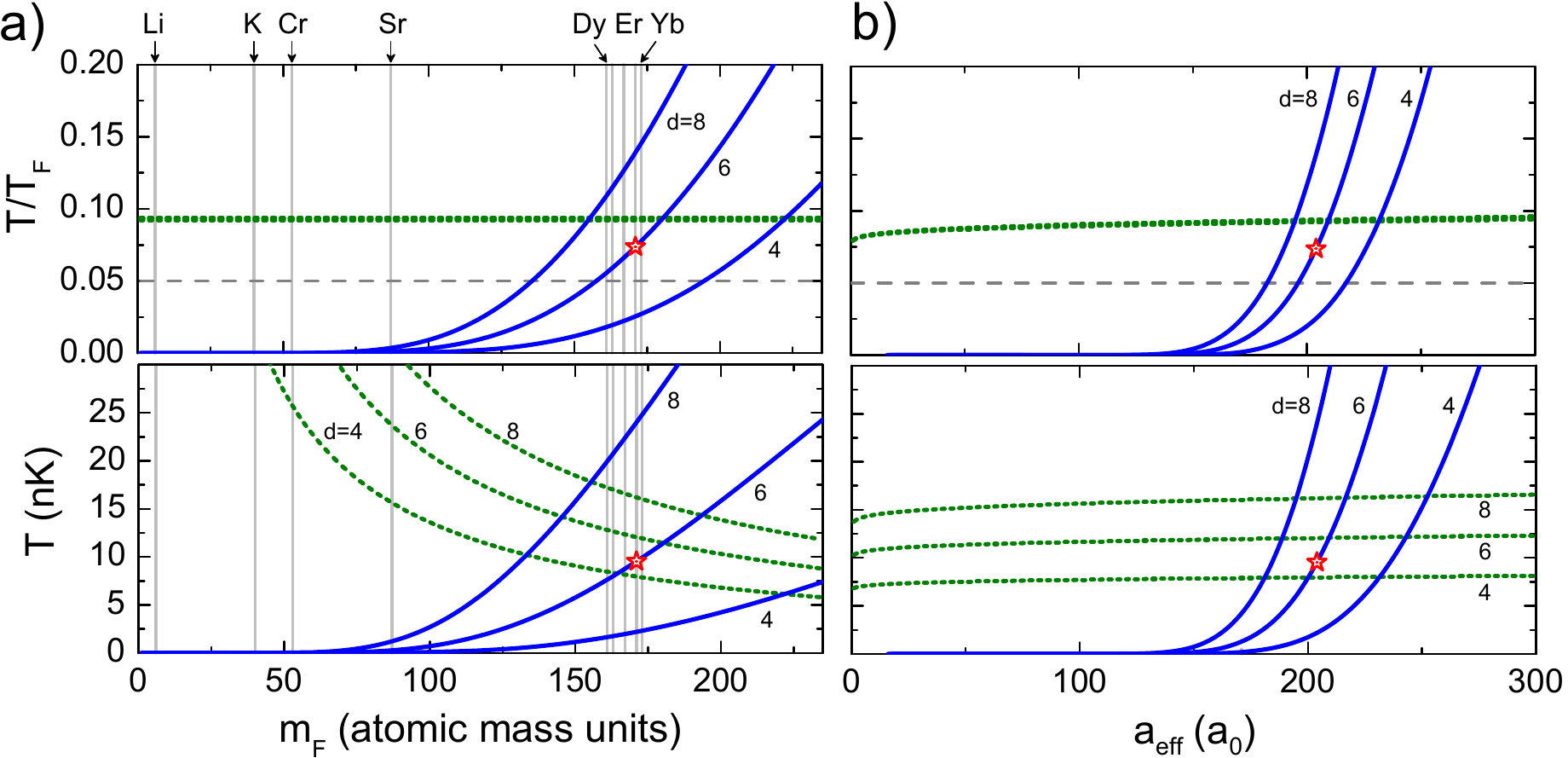}
  \caption{Maximum $p$-wave superfluid critical temperature $T_c^p/T_F$ (upper panels, solid lines) and $T_c^p$ (lower panels, solid lines) for fermions immersed in bosonic $^7$Li, as well as $T_{\rm KT}/T_F$ (upper panels, dotted lines) and $T_{\rm KT}$ (lower panels, dotted lines). a) Dependence on the mass of the fermions $m_F$. Here $n_B=d \times 10^{14}$\,atoms/cm$^3$, $a_B=8\,a_0$, $a_{\rm eff}=200\,a_0$ (corresponding to $\gamma_{BEC}=0.002\,d^{1/3}$ and $\gamma_{\rm eff}=0.05\,d^{1/3}$), and $\alpha=1.5\,\alpha_{\rm opt}$. Fermionic isotopes of elements that have been cooled to quantum degeneracy are marked by vertical lines. b) Dependence on $a_{\rm eff}$ for the fermion $^{171}$Yb, with all other parameters as before. The dashed lines in the upper panels mark the experimentally achieved $T/T_F$. The stars mark the example detailed in Table~\ref{TabLiYbParameters}.  }
  \label{fig1exp}
\end{figure*}

\begin{table}[!ht]
\centering
\caption{Parameters of $^{171,173}$Yb-$^7$Li mixture. The elastic scattering rate $\Gamma_{\rm el, B}$ is given for thermal atoms at a temperature of $T=T_c^p$ colliding with a BEC at density $n_B$. $\Gamma_{\rm 3-body, B}=-\dot{N}_B/N_B$ is the initial 3-body loss rate of the BEC \cite{Gross2009oou,Pollack2009uit}.}
\label{TabLiYbParameters} 
\begin{tabular}{@{\extracolsep{\fill}}ll}
\hline\hline \noalign{\smallskip}
$n_B$ & $6\times 10^{14}$\,atoms/cm$^3$ \\
$a_B$ &  $8\,a_0$ \\
$a_{\rm FB}$ & $200\,a_0$ \\
$a_{\rm eff}$ & $\sqrt{m_B/m_{FB}}\,a_{\rm FB}=204\,a_0$ \\
$\alpha$ &  $v_F/c_s=1.5\,\alpha_{\rm max}=0.22$ \\
$\gamma_{BEC}$ &  $a_B n_B^{1/3}=0.004$ \\
$\gamma_{\rm eff}$ &  $a_{\rm eff} n_B^{1/3}=0.1$ \\
$\xi$ & $1/\sqrt{8\pi n_B a_B}=0.4\,\mu$m\\
$X$ & $\xi k_F = \xi \sqrt{4\pi n_F} = 3.8$ \\
$v_F$ & $\hbar k_F/m_F=0.4\,$cm/s\\
$c_s$ & $\sqrt{n_B g_B/m_B}=1.6\,$cm/s\\
$\Gamma_{\rm el, B}$ & $21\,$s$^{-1}$ \\
$\Gamma_{\rm 3-body, B}$ & $0.002\,$s$^{-1}$ \\
$\mu_{BEC}$ &  $g_B n_B = k_B\times 221\,$nK$=h\times 4.6\,$kHz \\
$T_c^{\rm BEC}$ & 16.4\,$\mu$K \\
$n_F$ &  $720\,$atoms/(10\,$\mu$m)$^2$ \\
$E_F$ & $k_B\times 130\,$nK$=h\times 2.7\,$kHz$=0.6\,\mu_{BEC}$ \\
$T_c^p$ & $0.07\,T_F=5\times 10^{-4}\,T_c^{\rm BEC}=9.5\,$nK \\
$T_{\rm KT}$ & $0.09\,T_F=12\,$nK \\
\hline \hline
\end{tabular}
\end{table}

We now discuss possible choices of elements for the mixture. Since $m_F/m_B$ should be large, we limit our choice of bosons to the lightweight isotopes that have been Bose condensed, $^4$He$^*$, $^7$Li, and $^{23}$Na. Among those, $^7$Li has the great advantage of possessing a broad Feshbach resonance, with which $a_B$ can be tuned \cite{Chin2010fri,Pollack2009eto,Gross2009oou,Pollack2009uit}. Feshbach resonances in $^4$He$^*$ and $^{23}$Na are expected or known to be accompanied by strong losses \cite{Vassen2012cat,Goosen2010fri,Borbely2012mfd,Inouye1998oof,Stenger1999sei}. In the following, we use the triplet-scattering length for $^4$He$^*$ and $^{23}$Na \cite{Tiesinga1996asd,Moal2006ado}. Considering BEC densities for which inelastic collisions limit the BEC lifetime to 10\,s \cite{Robert2001abe,PereiraDosSantos2001bec,StamperKurn1998oco}, fermion masses up to the mass of the heaviest naturally occurring fermionic isotope ($^{235}$U) and $a_{\rm eff}=600\,a_0$, we find that $T_c^p/T_F<10^{-2}$ for these bosons. Only larger values of $a_{\rm eff}$ might make them suitable for our purposes.

We therefore limit our considerations to $^7$Li. This choice makes it possible to decrease $a_B$ and thereby increase $T_c^p/T_F$. To {choose} the fermionic element we plot in Fig.\,\ref{fig1exp}a) $T_c^p/T_F$ and $T_c^p$ as a fuction of $m_F$. Fermionic isotopes that have been cooled to quantum degeneracy and for which the experimentally relevant regime $T_c^p/T_F>0.05$ can be reached are $^{171,173}$Yb, $^{161}$Dy, and $^{167}$Er \cite{DeMarco1999oof,Naylor2015cdf,DeSalvo2010dfg,Taie2010roa,Lu2012qdd,Aikawa2014rfd}. A drawback of having to choose such heavy elements could be that they are not well sympathetically cooled by the lightweight Li because during each elastic collision, the energy transfer from the fermion to the boson is suppressed by $4 m_F m_B/(m_F+m_B)^2 \sim 0.15$ \cite{Mudrich2002scw}. A benefit of Dy and Er compared to Yb is that several interspecies Feshbach resonances will likely be available across the broad $^7$Li Feshbach resonance, making it possible to tune $a_{B}$ and $a_{\rm FB}$ somewhat independently and to access large values of $a_{\rm FB}$, which would also make tuning of $a_{\rm eff}$ by confinement induced resonances possible.

Nevertheless, since $^{173,174}$Yb-$^6$Li mixtures are already available in the lab \cite{Hara2011qdm,Hansen2013poq}, we concentrate our discussion now on $^{171,173}$Yb-$^7$Li. Adapting the existing machines to operate with $^7$Li instead of $^6$Li should be straightforward. There are two fermionic Yb isotopes readily available, each providing a chance of possessing suitable interspecies interaction properties with $^7$Li. Figure\,\ref{fig1exp}b) shows the dependence of $T_c^p/T_F$ and $T_c^p$ on $n_B$ and $a_{\rm eff}$. Choosing $a_B=8\,a_0$ leads to the system parameters given in Table~\ref{TabLiYbParameters}. The dotted lines in Fig.\,\ref{fig1exp} are an estimation of the Kosterlitz-Thouless transition temperature, which is given by \cite{Fisher,Prokofev2001tdw} \begin{equation}\label{TKT2}
T_{\rm KT} = 4 \pi \frac{\hbar^2}{2 m} n \ln^{-1}\Big[\ln\Big(\frac{1}{na^2}\Big)\Big],
\end{equation} where $m$ and $n$ are the mass and density of the superfluid species, while $a$ characterizes the range of the interaction. In particular, for our case of fermionic-pair formation, the interaction between fermions that will form the Cooper pairs is proportional to $a_{FB}^2$, with  $m=2m_F$ and $n\sim n_F/2$. Eq.~(\ref{TKT2}) is valid for small interaction parameters $a_B$ and $a_{FB}$ - the first makes the range of the potential long enough, such that the superfluid fraction achieves its maximum value \cite{Nishida2010poa,bruun}. 

The critical temperature $T_c^p = 0.07\,T_F = 9.5\,$nK is in the regime of temperatures that have already been achieved experimentally, albeit in systems with larger elastic scattering length. However, $T_c^p/T_c^{\rm BEC} = 5 \times 10^{-4} $ is more than one order of magnitude lower than what has been reached so far. To enhance evaporative cooling, it might be useful to first evaporate at a scattering length above 100\,$a_0$ and to tune the scattering length to a lower value only when approaching the required low temperature, while compressing the gas at the same time. In doing so, one could even profit from a Li 3-body recombination minimum at $a_B=119\,a_0$ \cite{Pollack2009uit}.

 \begin{figure*}[!ht]
  \centering
  \includegraphics[width=\textwidth]{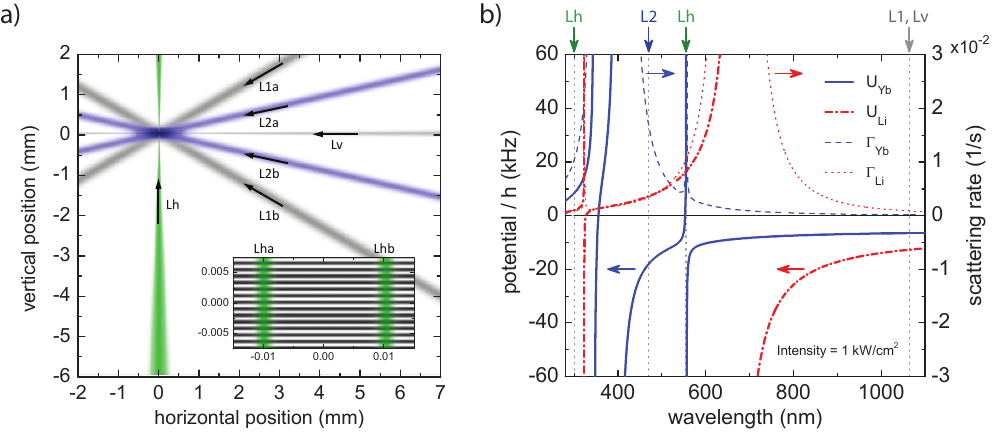}
  \caption{Mixed-dimensional optical dipole trap. a) Beam configuration. Ytterbium is confined in a 2D plane of an optical lattice formed by two standing waves created by laser beam pairs L1a,b and L2a,b. Both standing waves have the same intensity profile near the trap centre and are attractive for Yb, but generate opposite potentials for Li. Lithium is confined vertically by an elliptical Gaussian beam (Lv), elongated in the out-of-plane direction. Both elements are horizontally confined by four repulsive dipole-trap walls (Lha,b,c,d), forming a rectangular box. The inset shows the region around the trap centre, with Lha,b in cross section and the lattice intensity profile. b) Dipole potential and scattering rate for Li and Yb, as a function of the wavelength \cite{Grimm2000odt,NISTDatabase}. The arrows above the graph indicate the wavelengths of the dipole-trap beams. Two choices are possible for Lh.}
  \label{fig2exp}
 \end{figure*}

\subsection{Trap configuration}

Next, we consider suitable trap configurations for the mixture. Whereas the bosons explore a 3D trap, the fermions have to be effectively confined in 2D by a harmonic trap of frequency $\nu_{\perp, F}$, which requires $h \nu_{\perp, F}-E_F \gg k_B T$. The sample should be as homogeneous as possible to avoid inhomogeneous broadening of $p$-wave superfluidity signals, especially because the number of fermions will be low. Efficient evaporative cooling of the bosons should be possible in order to reach low temperatures. We now take these requirements into account to design an optical dipole trap for the mixture, where we orient the 2D plane of the fermions in the horizontal direction, see Fig.~\ref{fig2exp}a.

The bosonic lithium surrounds the fermions and can be confined by a Gauss-beam dipole trap using a wavelength of 1064\,nm. To reach a temperature $T$ by evaporation, the trap depth in the vertical direction $U_{\perp, B}$ should be $\mu_{\rm BEC} + \eta k_B T$, where $\mu_{\rm BEC}$ is the chemical potential of the BEC, and $\eta \sim 5$ {\cite{Ketterle1996eco}}. In order to provide a homogeneous vertical trap frequency across the cloud, the horizontal waist should be much larger than the cloud and the vertical Rayleigh length $z_R$ much longer than the horizontal sample size. The latter requirement and the additional requirement $h \nu_{\perp, B} \ll \mu_{\rm BEC}$ are only fulfilled if the vertical waist is larger than a minimum size. At the same time, the vertical waist should not be too large in order to limit the size of the $^7$Li sample in the vertical direction, thereby reducing the required number of $^7$Li atoms. Gravitational sag of the bosonic cloud is compensated by placing the focus of the Gauss beam slightly above the plane of the fermions. The Gaussian-beam trap creates a nearly constant potential on the fermions, since they explore only a small region in the centre of the trap. A constant potential offset is irrelevant and we can therefore ignore the influence of the Gauss-beam dipole trap on the fermions.

\begin{figure}[!ht]
  \centering
  \includegraphics[width=0.48 \textwidth]{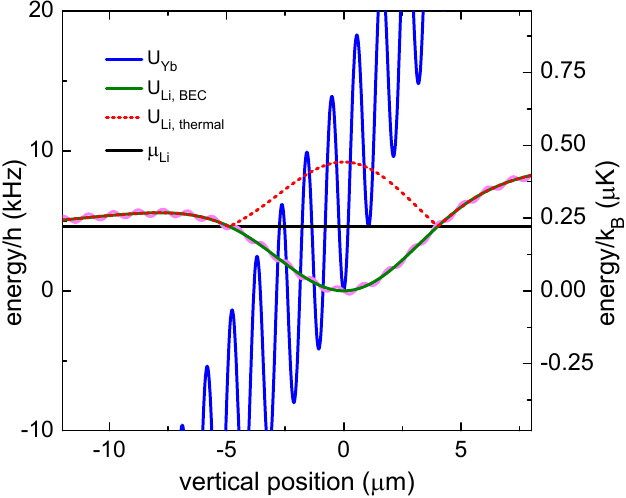}
  \caption{Optical dipole trap potential. A lattice confines Yb in 2D, whereas Li is levitated against gravity by a Gaussian beam. The potential experienced by thermal atoms $U_{\rm Li, thermal}$ consists of the dipole potential and twice the BEC mean-field potential \cite{PethickBook}. A phase fluctuation of a lattice beam by 0.1\,rad leads to the modulated Li potential shown around the ideal potential.}
  \label{fig3exp}
\end{figure}

To provide homogeneous confinement for bosons and fermions in the horizontal plane, repulsive dipole trap walls can be erected around the sample using vertically propagating Gauss beams \cite{Meyrath2005bec,Gaunt2013bec}. Four such beams can form a rectangular box with a size of $\sim 10\,\mu$m around the sample, if the waist of the beams is elongated along the sides of the rectangle ($w_{\rm Lh,\parallel}$ of a few 10\,$\mu$m) and is narrow orthogonal to that direction ($w_{\rm Lh,\perp}\sim 2\,\mu$m). This rectangular potential box also serves to select the most homogeneous central region of the traps that are used to confine bosons and fermions vertically. The sample density can easily be changed by moving the vertical walls towards each other, which is useful to do while $a_B$ is reduced to a low value. If in further studies a cylindrically symmetric system is required, for example to enable the creation of vortices \cite{Madison2000vfi}, a Laguerre-Gaussian beam can be used to confine the atoms horizontally \cite{Kaplan2002osb,Jaouadi2010bec,Gaunt2013bec}.

The confinement of the fermions in quasi-2D is most conveniently done using optical lattices. In comparison to other trap configurations, such as a Hermite-Gaussian beam \cite{Meyrath2005ahf,Meyrath2005bec}, it is easier to create a more homogeneous confinement in the 2D plane by increasing the diameter of the lattice beams. In order to populate only a single plane of the lattice with fermions, one can use the techniques of Refs. \cite{Gemelke2009iso, Sherson2010sar, Yamamoto2016ayq, Ville2016lac}.

The deep dipole potential used to confine the fermions in 2D may only have a negligible effect on the bosons. The parasitic potential on the bosons $U_{\rm lattice, B}$ must be much smaller than $\mu_{\rm BEC}$. This challenge has been met by species-specific dipole traps using a "tune-out" wavelength, for which the AC polarizability of one species is zero \cite{Massignan2006tds,LeBlanc2007sso,Catani2009eei,Lamporesi2010sim}. Unfortunately, this technique does not work for $^7$Li because its "tune-out" wavelength is too close to an atomic transition, leading to detrimental off-resonant scattering for the required trap depths \cite{LeBlanc2007sso}. Another option is to use a "tune-in" wavelength, close to an Yb transition and far detuned from any Li transition \cite{LeBlanc2007sso}. In this situation, the potential on Yb $U_{\rm lattice, F}$ can exceed the potential on Li many times. This technique is suitable for our situation, but will limit the lifetime of the fermionic cloud to a few seconds by off-resonant scattering. If this limit is significant depends on the other factors limiting the lifetime of the system, especially the unknown 3-body loss rate $\Gamma_{\rm FBB}$.

\begin{table}[!ht]
	\centering
\caption{Optical dipole trap configuration. $\lambda_{{\rm L}i}$ is the wavelength of dipole-trap beam ${\rm L}i$, with $i=1,2$. $w$ are the $1/e$ beam radii. The vertical trap depth for $^7$Li, $U_{\perp, B}$, takes the effect of gravity into account. $\alpha_{{\rm L}i}$ is the angle between lattice beams ${\rm L}i{\rm a}$ and ${\rm L}i{\rm b}$. $\Delta z$ is the lattice spacing. $n_{\rm 2D, B}$ is the density of bosons integrated over the vertical direction. $\tau_{\rm B,F}=1/\sum_i \Gamma_{i, {\rm B,F}}$ are {limits to }the lifetimes of bosons and fermions, where $\Gamma_{i, {\rm B,F}}$ is the off-resonant scattering rate of photons calculated at peak intensity of dipole trap beam L$i$, with $i$ running over all beams \cite{Grimm2000odt,McKay2011cis,Gordon1980moa}. }
	\label{TabDipoleTrapConfig} 
		\begin{tabularx}{0.45  \textwidth}{@{\extracolsep{\fill}}lXll}
\hline\hline \noalign{\smallskip}
$\lambda_{\rm Lv}$ &  1064\,nm  & $w_{\rm Lv}$ &  6\,$\mu$m   \\
$z_R$ & 100\,$\mu$m & & \\
$U_{\perp, B}$ & $k_B\times 0.27\,\mu$K   & $\nu_{\perp, B}$ & 1.1\,kHz   \\
$\lambda_{\rm Lh}$ &  300\,nm or 554\,nm   \\
$w_{\rm Lh,\perp}$ & 2\,$\mu$m & $w_{\rm Lh,\parallel}$ & 200\,$\mu$m   \\
$\lambda_{\rm L1}$ & 1064\,nm   &  $\alpha_{\rm L1}$ &  60\degree \\
$\lambda_{\rm L2}$  &  470\,nm & $\alpha_{\rm L2}$  &  25.5\degree   \\
$\Delta z$ &  1064\,nm    \\
$U_{\perp, F}$ & $h \times 16\,$kHz   & $\nu_{\perp, F}$ & 4.1\,kHz=1.5\,$E_F$   \\
$\tau_{\rm B}$ & 296\,s  & $\tau_{\rm F}$ & 79\,s  \\
$n_{\rm 2D, B}$ & \multicolumn{2}{l}{$3\times 10^5\,$atoms/(10\,$\mu$m)$^2$} &\\
\hline \hline
\end{tabularx}
\end{table}

If the lifetime limit imposed by a "tune-in" lattice is too severe, a bichromatic dipole trap can be used, consisting of two optical lattices that both confine Yb, but compensate each other for Li. This technique overcomes the possibly excessive off-resonant scattering and replaces it by the technical challenge of creating two lattices with very well controlled intensity profiles. We will explore this scheme in the following. We chose optical lattices with wavelengths of 470\,nm and 1064\,nm, which are both attractive for Yb. In contrast, for Li only the 1064-nm lattice is attractive, the other is repulsive, see Fig.\,\ref{fig2exp}b. In order for the lattice potentials to add up for Yb and cancel sufficiently for Li, the intensity profile of both lattices need to be nearly identical in the region of the atomic clouds. The lattice-well spacing must be the same, and the intensity maxima need to overlap. The lattice spacing can be adjusted by the angle between the two lattice beams of each wavelength. Using an angle of 60$^\circ$ between the two beams forming the 1064-nm lattice leads to a lattice spacing of 1064\,nm. The same spacing is reached for the 470-nm lattice if the two corresponding beams intersect at an angle of 25.5$^\circ$, see Fig.\,\ref{fig2exp}a. The position of the intensity maxima along the lattice direction (the vertical direction) depends on the phase difference between the two beams forming a lattice. This phase difference has to be stabilized interferometrically for each lattice to a common reference, combining methods from Refs.\,\cite{Foelling2007doo,Wirth2010efo}. In order for the two lattice potentials to cancel for the bosons, the intensity of the 470-nm lattice beams has to be 1.8 times the intensity of the 1064-nm lattice beams. For Yb the two lattice potentials add up, giving a total potential that is 1.2 times larger than the potential of the 470-nm lattice alone. This total potential needs to confine Yb in quasi-2D and be also deep enough to suppress tunneling of Yb to neighboring lattice planes, see Fig.\,\ref{fig3exp}. The cancelation of the lattice potential for the bosons will not be perfect because of intensity and phase fluctuations leading to deviations from the ideal configuration. Phase fluctuations of 90\,mrad or intensity imbalances of 9\% lead to a residual potential on the order of 10\% of $\mu_{\rm BEC}$. This parasitic potential would be tolerable if the timescale of fluctuations is large enough to avoid heating of the sample. In principle, we could have chosen a wavelength for L2 that is further away from the Yb transition, e.g. 532\,nm, which would reduce off-resonant scattering and simplify phase locking of the laser sources used for L1 and L2. All the same, we chose 470-nm because at that wavelength we are profiting from less parasitic potential of L2 on Li, reducing the amount of compensation needed from L1. As a result, the overall parasitic potential created for a given intensity or phase mismatch between L1 and L2 is reduced.

Example parameters for the bichromatic dipole trap and important results of using this trap for the Li-Yb mixture are given in Table~\ref{TabDipoleTrapConfig}. The $^7$Li atom number available in current experiments ($3\times 10^5$ atoms \cite{Pollack2009eto}) is sufficient for a square sample of 10\,$\mu$m size. A sample of this size contains about 700 fermions. If this proposal is realisable depends to a large extend on the unknown elastic and inelastic scattering properties of Li-Yb. Similar schemes can be applied to other mixtures, such as Li-Dy or Li-Er, for which some interspecies interaction tuning should be possible.

\subsection{Detection of $p$-wave superfluidity}
There are some predictable signatures for the experimental detection of the $p_{x} + ip_{y}$ superfluid phase. Particularly, the density of state (rf absorption spectrum) of a rotating weak pairing  $p_{x} + ip_{y}$ phase is expected to exhibit a set of gapless modes \cite{Grosfeld2007pso}, which are a direct consequence of the zero-energy Majorana modes on the vortices. The rf-spectroscopy can be also applied to detect Majorana edge states of the topological superfluid in a 2D square lattice \cite{Midtgaard2016tso}.
On the other hand,  the time-reversal symmetry broken signature of the chiral $p_{x} + ip_{y}$ fermionic superfluid can be detected with time-of-flight image of the atomic density distribution: an external effective electric field (i.e., dipole interaction between the neutral atoms in the superfluid and the laser field) brings a nonzero antisymmetric transverse mass current in the velocity distribution of the atoms \cite{Zhang2008pxp}.

\section{Conclusion}
In the present work, we explored the feasibility of a $p$-wave superfluid by using a Fermi-Bose mixture in a mixed-dimension configuration, where $p$-wave interaction between spin-polarized degenerate fermions in 2D is induced indirectly, through the scattering of the Bogoliubov modes of condensed bosons moving in 3D. We have shown that, even in the weak-coupling regime, the appropriate renormalization of the phonon propagator (BEC modes) with particle-hole fluctuations and the vertex correction significantly increase the gap and the predicted critical temperature for the fermion-pair formation.

It is important to remark that we adopt a minimum value for $\gamma_{BEC}\sim a_B n_B^{1/3}$, which yields $\upsilon_F/c_s \leq 1$, thus allowing to disregard retardation effects. According to Wu and Bruun \cite{bruun}, who performed calculations including retardation but no vertex correction to determine $T_{MF}$, in the limit $\upsilon_F/c_s \leq 1$, it holds that $T_{MF} \sim T_{BCS}$ (see Fig.2 in the cited reference), which confirms the validity of our approximation.

We neglected decay of the BEC phonons, like the Beliaev damping and the lifetime due to the scattered particle-hole pairs of the degenerate fermionic sample. The Beliaev damping is given by the boson-boson scattering potential, resulting in a phonon lifetime proportional to $g_B$ \cite{Davidson, Shlyapnikov2}. In the small-momentum regime, however, the Beliaev decay mechanism is strongly suppressed \cite{Davidson}. On the other hand, if we consider the phonon dressed by particle-hole fluctuations of the Fermi sea, it will have a lifetime proportional to $g_{FB}^2$. In the static limit considered in the paper, however, the lifetime is infinite (see App.~\ref{apB} for details). Hence, we conclude that there is no damping mechanism that could hamper the stability of the BEC in the chosen regime of parameters.

Exploiting the difference in polarizability and mass of the atomic species, and by optimizing the density $n_B$ and the scattering length $a_B$ of the bosons, our work sets the boundary for the experimental realization of a $p$-wave superfluid within the reachable limit of $T^p_c = 0.05 T_F$. It identifies a realistic route and provides the details to the accomplishment and manipulation of this long-sought fascinating chiral-superfluid phase in the realm of ultracold atoms in optical lattices.

\section*{Acknowledgments}
We thank Rodrigo G. Pereira, Servaas Kokkelmans, Fr\'{e}d\'{e}ric Chevy, and Subhadeep Gupta for discussions and insightful comments. This work was supported by CNPq (Brazil) through the Brazilian government project Science Without Borders. The work of C.M.S. is part of the DITP consortium, a program of the Netherlands Organisation for Scientific Research (NWO) that is funded by the Dutch Ministry of Education, Culture and Science (OCW). {F.S. gratefully acknowledges funding from the European Research Council (ERC) under Project No. 615117 QuantStro and by NWO through Vici grant No. 680-47-619.}\\

\bibliographystyle{apsrev4-1}

\bibliography{FermiBoseRef}

\begin{thebibliography}{82}
\bibitem [{cal()}]{calculo}%
  \BibitemOpen
  \href@noop {} {}\bibinfo {note} {We considered the Fermi-Bose mixtures
  $^{87}$Sr-$^{7}$Li and $^{171}$Yb-$^{7}$Li, assuming $a_B\sim 20 a_0$ and
  with the density of $^{7}$Li limited to $n_B \sim 10^{14}$
  atoms/cm$^3$.}\BibitemShut {Stop}%
\end{thebibliography}

\newpage

\onecolumngrid

\appendix

\section{Bogoliubov transformation in the BCS Hamiltonian} \label{apA}

Starting with the definition \begin{eqnarray}
  \hat{A}(\mathbf{k},\mathbf{p}) &=& \hat{a}\left({\mathbf{k}}/{2}-\mathbf{p}\right)\hat{a}\left({\mathbf{k}}/{2}+\mathbf{p}\right),
\end{eqnarray} we can apply a mean-field approach in Eq.~(\ref{eq4}) and replace the pair operator $\hat{A}(\mathbf{k},\mathbf{p})$ by $\langle\hat{A}(\mathbf{k},\mathbf{p})\rangle  +
\delta\hat{A}(\mathbf{k},\mathbf{p})$ (similar expression for its conjugate), with $\langle\hat{A}(\mathbf{k},\mathbf{p})\rangle = \delta_{\mathbf{k},\mathbf{0}}
\langle\hat{a}(-\mathbf{p})\hat{a}(\mathbf{p})\rangle$ and $\langle\hat{A}^{\dagger}(\mathbf{k},\mathbf{p})\rangle = \delta_{\mathbf{k},\mathbf{0}}
\langle\hat{a}^{\dagger}\big(\mathbf{p}\big)\hat{a}^{\dagger}(-\mathbf{p})\rangle$. Holding terms up to the first order in the fluctuations of this field (neglecting $\mathcal{O}[(\delta\hat{A})^n], \;
n>1)$, we find \begin{eqnarray} \hat{H}_F^{BCS} = \int \frac{d^2p}{(2\pi)^2} \bigg\{ \epsilon_{p} \hat{a}^{\dag}(\mathbf{p})\hat{a}(\mathbf{p}) + \frac{1}{2}
{\triangle}^{\ast}_{\mathbf{p}} \big\langle\hat{a}(-\mathbf{p})\hat{a}(\mathbf{p}) \big\rangle  -\frac{1}{2} \left[{\triangle}^{\ast}_{\mathbf{p}}
\hat{a}(-\mathbf{p})\hat{a}(\mathbf{p})+{\triangle}_{\mathbf{p}} \hat{a}^{\dagger}(\mathbf{p})\hat{a}^{\dagger}(\mathbf{-p})\right]\bigg\},
\end{eqnarray} with $\epsilon_{p}  =  {p^2}/{2m_F}-\mu$ and the order parameter (or momentum-dependent gap) expressed as \begin{eqnarray} \label{eqgap}
\triangle_{\mathbf{p}} = - \int \frac{d^2k}{(2\pi)^2} V_{\textrm{eff}}(\mathbf{p},\mathbf{k})  \big\langle\hat{a}(-\mathbf{k})\hat{a}(\mathbf{k}) \big\rangle,
\end{eqnarray} where we consider the interaction potential  \begin{eqnarray} \label{veff}
V_{\textrm{eff}}(\mathbf{p},\mathbf{k}) = - V_0\frac{1}{\sqrt{|\mathbf{p}-\mathbf{k}|^2+2\xi^{-2}}}, \end{eqnarray} with $V_0 = 2 g_{FB}^2n_Bm_B$. Before applying the Bogoliubov transformation, let us
first symmetrize this BCS Hamiltonian properly. It is easier to go further with this process in the discrete-momentum space, summing over half of the k-space $\sum_{\mathbf{k}} \rightarrow
\sum_{\mathbf{k}}^{\prime}$ \begin{eqnarray}  \nonumber  \hat{H}_F^{BCS}  && =  \sum_{\mathbf{p}} \bigg[\epsilon_{p}
\hat{a}^{\dag}_{\mathbf{p}}\hat{a}_{\mathbf{p}}-\frac{1}{2}\left({\triangle}^{\ast}_{\mathbf{p}} \hat{a}_{-\mathbf{p}}\hat{a}_{\mathbf{p}}+{\triangle}_{\mathbf{p}}
\hat{a}^{\dagger}_{\mathbf{p}}\hat{a}^{\dagger}_{\mathbf{-p}}\right)  + \frac{1}{2} {\triangle}^{\ast}_{\mathbf{p}} \big\langle\hat{a}_{-\mathbf{p}}\hat{a}_{\mathbf{p}} \big\rangle
\bigg] \\
&&= {\sum_{\mathbf{p}}}^{\prime} \bigg[ \epsilon_{p} \left(\hat{a}^{\dag}_{\mathbf{p}}\hat{a}_{\mathbf{p}}+\hat{a}^{\dag}_{\mathbf{-p}}\hat{a}_{\mathbf{-p}}\right)-
\left(\hspace{-0.1cm}{\triangle}^{\ast}_{\mathbf{p}} \hat{a}_{-\mathbf{p}}\hat{a}_{\mathbf{p}}+{\triangle}_{\mathbf{p}} \hat{a}^{\dagger}_{\mathbf{p}}\hat{a}^{\dagger}_{\mathbf{-p}}\right)  +
{\triangle}^{\ast}_{\mathbf{p}} \big\langle\hat{a}_{-\mathbf{p}}\hat{a}_{\mathbf{p}} \big\rangle \bigg],
\end{eqnarray} where we used the property ${\triangle}_{-\mathbf{p}}=-{\triangle}_{\mathbf{p}}$, which is simple to prove if we consider that
$V_{\textrm{eff}}(-\mathbf{p},\mathbf{k})=V_{\textrm{eff}}(\mathbf{p},-\mathbf{k})$ and $V_{\textrm{eff}}(-\mathbf{p},-\mathbf{k})=V_{\textrm{eff}}(\mathbf{p},\mathbf{k})$, as can be promptly verified
from Eq.~(\ref{veff}).

Now, we apply the canonical transformation  \begin{eqnarray}  \nonumber
 \hat{a}_{\mathbf{p}} = u_{\mathbf{p}} \hat{\alpha}_{\mathbf{p}} + v_{\mathbf{p}} \hat{\alpha}^{\dagger}_{-\mathbf{p}} \\
 \hat{a}^{\dagger}_{-\mathbf{p}}= -\bar{ v}_{\mathbf{p}} \hat{\alpha}_{\mathbf{p}} + \bar{u}_{\mathbf{p}} \hat{\alpha}^{\dagger}_{-\mathbf{p}},
 \end{eqnarray} with $|u_{\mathbf{p}}|^2+|v_{\mathbf{p}}|^2=1$. To diagonalize the transformed Hamiltonian, we set the coefficients of the off-diagonal terms to zero, $2
 \epsilon_{p}u_{\mathbf{p}}\bar{v}_{\mathbf{p}}-{\triangle}^{\ast}_{\mathbf{p}}u_{\mathbf{p}}^2+{\triangle}_{\mathbf{p}}\bar{v}_{\mathbf{p}}^2=0$. Multiplying this equation by
 ${\triangle}_{\mathbf{p}}/u_{\mathbf{p}}^2$, we get \begin{eqnarray}
 2 \epsilon_{p} \frac{{\triangle}_{\mathbf{p}}\bar{v}_{\mathbf{p}}}{u_{\mathbf{p}}}
 -|{\triangle}_{\mathbf{p}}|^2
 +\frac{{\triangle}_{\mathbf{p}}^2\bar{v}_{\mathbf{p}}^2}{u_{\mathbf{p}}^2}=0,
 \end{eqnarray} and then
 \begin{eqnarray} \label{eqconj}
 \frac{{\triangle}_{\mathbf{p}}\bar{v}_{\mathbf{p}}}{u_{\mathbf{p}}}=E_{p}-\epsilon_{p},
 \end{eqnarray} with the energy dispersion $E_{p}=\sqrt{\epsilon_{p}^2+|{\triangle}_{\mathbf{p}}|^2}$. Using the conjugate of Eq.~(\ref{eqconj}), we can prove that
 $\frac{|v_{\mathbf{p}}||{\triangle}_{\mathbf{p}}|}{|u_{\mathbf{p}}|}=E_{p}-\epsilon_{p}$. Now, with the previous relation for the parameters $u_{\mathbf{p}}$ and $v_{\mathbf{p}}$, we find
 \begin{eqnarray}
|u_{\mathbf{p}}|^2 &&= 1-|v_{\mathbf{p}}|^2 = \frac{1}{2}\left[1+\frac{\epsilon_{p}}{E_{p}}\right].
 \end{eqnarray}

Finally, we can build the diagonal form \begin{eqnarray}  \hat{H}_F^{BCS}
= {\sum_{\mathbf{p}}}^{\prime}   E_{p} \left(\hat{\alpha}^{\dag}_{\mathbf{p}}\hat{\alpha}_{\mathbf{p}}+\hat{\alpha}^{\dag}_{\mathbf{-p}}\hat{\alpha}_{\mathbf{-p}}\right) +{\sum_{\mathbf{p}}}^{\prime} \left[{\triangle}^{\ast}_{\mathbf{p}} \big\langle \hat{a}_{-\mathbf{p}}\hat{a}_{\mathbf{p}} \big\rangle + \left(\epsilon_{p}-E_{p}\right)\right].
\end{eqnarray} \\ Considering $\big\langle\hat{a}_{-\mathbf{p}}\hat{a}_{\mathbf{p}} \rangle =
-u_{\mathbf{p}}v_{\mathbf{p}}\big\langle\hat{\alpha}^{\dag}_{\mathbf{p}}\hat{\alpha}_{\mathbf{p}}\big\rangle+u_{\mathbf{p}}v_{\mathbf{p}}\langle\hat{\alpha}_{-\mathbf{p}}\hat{\alpha}^{\dag}_{-\mathbf{p}}\big\rangle$,
with $\big\langle\hat{\alpha}^{\dag}_{\mathbf{p}}\hat{\alpha}_{\mathbf{p}}\big\rangle = n_{F}(E_{p}) = [\exp(\beta E_{p}) +1]^{-1}$, where $\beta=(k_BT)^{-1}$, we obtain the final result
 \begin{eqnarray}  \hat{H}_F^{BCS}
 = {\sum_{\mathbf{p}}}   E_{p} \hat{\alpha}^{\dag}_{\mathbf{p}}\hat{\alpha}_{\mathbf{p}} +
 \frac{1}{2}{\sum_{\mathbf{p}}}\bigg\{\frac{|{\triangle}_{\mathbf{p}}|^2}{E_{p}}\Big[1-2n_F(E_{p})\Big]+ \left(\epsilon_{p}-E_{p}\right)\bigg\}.\end{eqnarray}

\section{Higher-order correction to the effective 2D-3D interaction} \label{apB}
Starting with the interaction between the fermions in 2D and the ``phonons" of the BEC in 3D (see the main text) \begin{eqnarray}
\hat H_{int}(\tau)=g_{FB} \sqrt{n_B}\frac{1}{\sqrt{V}}\sum_{\mathbf{p}_1,\mathbf{p}_2,q_z} \; V_{q} \left[ \hat{\beta}_{\mathbf{q}}(\tau)+\hat{\beta}^{\dagger}_{-\mathbf{q}}(\tau)\right]
\hat{a}^{\dagger}_{\mathbf{p}_1}(\tau)\hat{a}_{\mathbf{p}_2}(\tau),
\end{eqnarray} where $\mathbf{q}\equiv(\mathbf{p}_1-\mathbf{p}_2,q_z)$ and \begin{eqnarray}V_{q}  = \bigg(\frac{q^2}{q^2+2\xi^{-2}}\bigg)^{1/4}.\end{eqnarray}
\begin{figure}[!ht]
  \centering
  \includegraphics[width=16.0cm]{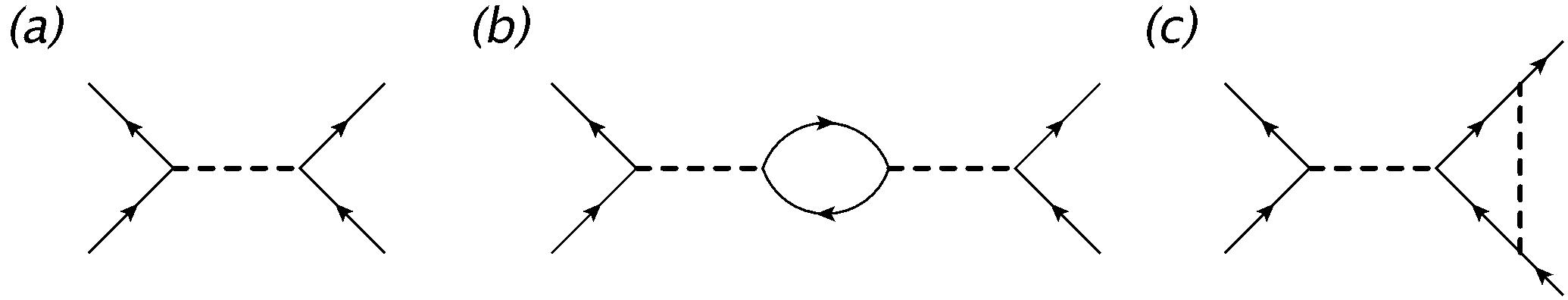}
  \caption{ \label{fig4} Second (a) and fourth-order, (b) and (c), Feynman diagrams for the effective interaction between two fermions in 2D.}
\end{figure}  \\ In the following we calculate the four-point function to $4th$ order in the interaction constant $g_{FB}$
\begin{eqnarray}
\Gamma(\{\mathbf{k}_i,\tau_i\})&&=- \bigg\langle T_{\tau} \hat{a}_{\mathbf{k}_1}(\tau_1)\hat{a}_{\mathbf{k}_2}(\tau_2)\hat{a}^{\dagger}_{\mathbf{k}_3}(\tau_3)\hat{a}^{\dagger}_{\mathbf{k}_4}(\tau_4)
e^{-\int_0^{\beta}d\tau \hat H_{int}(\tau)} \bigg\rangle,
\end{eqnarray} which corresponds to the Feynman diagrams shown in Fig.~\ref{fig4}. We consider the effective interaction between the fermions in 2D, with the free-fermion propagator given by
$\mathcal{G}_0$ \begin{eqnarray}
\Gamma_{\textrm{eff}}(\{\mathbf{k}_i,\nu_i\})=\lambda_{\textrm{eff}}\frac{1}{S}\delta_{\mathbf{k}_1+\mathbf{k}_2,\mathbf{k}_3+\mathbf{k}_4}\beta \delta_{\nu_1+\nu_2,\nu_3+\nu_4} \prod_{i=1...4}
\mathcal{G}_0(\mathbf{k}_i,\nu_i).\end{eqnarray} The second-order expansion in the coupling $g_{FB}$ provides [$\mathbf{q}\equiv(\mathbf{k}_1-\mathbf{k}_4,q_z)$] \begin{eqnarray} \nonumber
\Gamma^{(2)}(\{\mathbf{k}_i,\nu_i\})&&= \frac{1}{V} g_{FB}^2 n_B  \sum_{q_z} V^2_{\mathbf{q}} \mathcal{D}_0(\mathbf{q},\nu_1-\nu_4) \delta_{\mathbf{k}_1+\mathbf{k}_2,\mathbf{k}_3+\mathbf{k}_4}\beta
\delta_{\nu_1+\nu_2,\nu_3+\nu_4} \prod_{i=1...4} \mathcal{G}_0(\mathbf{k}_i,\nu_i) \\
&& = - 2 g_{FB}^2 n_B m_B \frac{1}{\sqrt{|\mathbf{k}_1-\mathbf{k}_4|^2+2\xi^{-2}}} \frac{1}{S} \delta_{\mathbf{k}_1+\mathbf{k}_2,\mathbf{k}_3+\mathbf{k}_4}\beta \delta_{\nu_1+\nu_2,\nu_3+\nu_4}
\prod_{i=1...4} \mathcal{G}_0(\mathbf{k}_i,\nu_i) ,\end{eqnarray} where we used static approximation to the Bogoliubov-mode propagator $\mathcal{D}_0$.

\subsection{RPA correction}
At higher-order expansion, we obtain for the diagram in Fig.~\ref{fig4}(b) \begin{eqnarray} \nonumber
&&\Gamma_{RPA}^{(4)}(\{\mathbf{k}_i,\nu_i\}) =
 \frac{g_{FB}^4 n_B^2 }{V^2} \delta_{\mathbf{k}_1+\mathbf{k}_2,\mathbf{k}_3+\mathbf{k}_4}\beta \delta_{\nu_1+\nu_2,\nu_3+\nu_4} \prod_{i=1...4} \mathcal{G}_0(\mathbf{k}_i,\nu_i) \times \\  && \times
 \sum_{\mathbf{p},q_{2z},q_{3z}} V^2_{\mathbf{q}_2}V^2_{\mathbf{q}_3} \mathcal{D}_0(\mathbf{q}_2,i\nu_2-i\nu_3)\mathcal{D}_0(\mathbf{q}_3,i\nu_4-i\nu_1) \sum_n
\mathcal{G}_0(\mathbf{p}+\mathbf{k}_4-\mathbf{k}_1,\nu_4-\nu_1+\nu_n)\mathcal{G}_0(\mathbf{p},\nu_n), \end{eqnarray} with $\mathbf{q}_2 = (\mathbf{k}_2-\mathbf{k}_3,q_{2z})$ and $\mathbf{q}_3 =
(\mathbf{k}_4-\mathbf{k}_1,q_{3z})$, which eventually leads to \begin{eqnarray} \nonumber  \label{eqRPA} \Gamma_{RPA}^{(4)}(\{\mathbf{k}_i,\nu_i\})=   \frac{4g_{FB}^4 n_B^2
m_B^2}{|\mathbf{k}_1-\mathbf{k}_4|^2+2\xi^{-2}} \frac{1}{S} \sum_{\mathbf{p}}
\frac{n_F(\epsilon_{\mathbf{p}})-n_F(\epsilon_{\mathbf{p}+\mathbf{k}_4-\mathbf{k}_1})}{\nu_4-\nu_1+\epsilon_{\mathbf{p}}-\epsilon_{\mathbf{p}+\mathbf{k}_4-\mathbf{k}_1}}\frac{1}{S}
\delta_{\mathbf{k}_1+\mathbf{k}_2,\mathbf{k}_3+\mathbf{k}_4}\beta \delta_{\nu_1+\nu_2,\nu_3+\nu_4} \prod_{i=1...4} \mathcal{G}_0(\mathbf{k}_i,\nu_i). \\ \end{eqnarray}

Now we will solve the ``polarization bubble'' in 2D \begin{eqnarray}
P(\mathbf{k},i\nu) = \int\frac{d^2p}{(2\pi)^2} \frac{n_F(\epsilon_{\mathbf{p}})-n_F(\epsilon_{\mathbf{p}+\mathbf{k}})}{i\nu+\epsilon_{\mathbf{p}}-\epsilon_{\mathbf{p}+\mathbf{k}}},
\end{eqnarray} Before we integrate in momentum space, we simplify the above expression by changing the variable in the second term to $\mathbf{p}^{\prime}=\mathbf{p}+\mathbf{k}$. We then obtain
\begin{eqnarray}\label{acima}
P(\mathbf{k},i\nu)= \int\frac{d^2p}{(2\pi)^2} n_F(\epsilon_{\mathbf{p}})
\bigg(\frac{1}{i\nu+\epsilon_{\mathbf{p}}-\epsilon_{\mathbf{p}+\mathbf{k}}}-\frac{1}{i\nu+\epsilon_{\mathbf{p}-\mathbf{k}}-\epsilon_{\mathbf{p}}}\bigg).\end{eqnarray} Since we are interested in the
zero-temperature limit, we consider the analytic continuation $i\nu \rightarrow \nu+i\delta$, with $n_F(\epsilon_{\mathbf{p}}) \rightarrow  \Theta(\mu-\varepsilon_{\mathbf{p}})$. Then, we focus on the real part of Eq.~(\ref{acima})  \begin{eqnarray} \label{eqCg}
\mathcal{R}e\:  P(\mathbf{k},\nu)= - \int_0^{k_F} \frac{p dp}{2\pi} \; \int_{-\pi}^{\pi} \frac{d\theta}{2\pi}\; \frac{2\varepsilon_{\mathbf{k}}}{\varepsilon_{\mathbf{k}}^2-\big(\frac{p k \cos\theta}{m_F}-\nu\big)^2}.
\end{eqnarray}

Starting with the angular integral in Eq.~(\ref{eqCg}) (for $|{k}/{2k_F}\pm{\nu m_F}/{k k_F}|>1$), after changing the variable $p \rightarrow \varepsilon = {p^2}/{2m_F}$ in the resulting integral, we
obtain (see Ref.\cite{PRL})
\begin{eqnarray}\nonumber \mathcal{R}e\: P(\mathbf{k},\nu) &&= - \frac{m_F}{2\pi}\int_0^{\mu} d\varepsilon
\bigg\{\frac{1}{\big[(\varepsilon_{\mathbf{k}}+\nu)^2-\frac{2k^2\varepsilon}{m_F}\big]^{1/2}}+\frac{1}{\big[(\varepsilon_{\mathbf{k}}-\nu)^2-\frac{2 k^2\varepsilon}{m_F}\big]^{1/2}}\bigg\} \\
&& =-\frac{m_F^2}{2\pi}\frac{1}{k^2}\bigg\{|\varepsilon_{\mathbf{k}}+\nu|+|\varepsilon_{\mathbf{k}}-\nu|-\sqrt{(\varepsilon_{\mathbf{k}}+\nu)^2-\frac{2 k^2
\mu}{m_F}}-\sqrt{(\varepsilon_{\mathbf{k}}-\nu)^2-\frac{2 k^2 \mu}{m_F}}\bigg\},
\end{eqnarray} remembering that $\mu={k_F^2}/{2m_F}$. Particularly, in the static limit $\nu = 0$, we will have \begin{eqnarray}
\mathcal{R}e\: P(\mathbf{k}) && = -  \frac{m_F}{2\pi} \qquad \textrm{for} \qquad    k < 2 k_F,
\end{eqnarray} and \begin{eqnarray} \mathcal{R}e\: P(\mathbf{k})
&& = - \frac{m_F}{2\pi}\bigg(1-\sqrt{1-\frac{4 k_F^2 }{{k}^2}}\bigg) \qquad \textrm{for} \qquad   k > 2 k_F.
\end{eqnarray} Assuming $|\mathbf{k}_1-\mathbf{k}_4| < 2 k_F $, we can easily calculated the RPA series, which gives \begin{eqnarray}  \nonumber
\lambda_{\textrm{eff}}^{RPA} &&= \lambda_0+\lambda_0^2 P_0+\lambda_0^3 P_0^2+ ... \\
&& =  \lambda_0 [ 1+ \lambda_0 P_0+\lambda_0^2 P_0^2+ ...],
\end{eqnarray} where we defined $\lambda_0 =  - {2 g_{FB}^2 n_B m_B}/{\sqrt{|\mathbf{k}_1-\mathbf{k}_4|^2+2\xi^{-2}}}$ and $P_0=-{m_F}/{2\pi}$. For $ \lambda_0 P_0<1$, we find \begin{eqnarray}
\lambda_{\textrm{eff}}^{RPA} = \frac{\lambda_0}{1-\lambda_0P_0} = - \frac{2 g_{FB}^2 n_B m_B}{\sqrt{|\mathbf{k}_1-\mathbf{k}_4|^2+2\xi^{-2}} - \frac{g_{FB}^2 n_B m_Bm_F}{\pi}}.\end{eqnarray}

Now, we consider the RPA correction to calculate the projected component $\ell=1$ of the potential $V_{\textrm{eff}}^{(1)}$, i.e.,\begin{eqnarray}
{\lambda_{\textrm{eff}}^{RPA}}^{(1)}(k_{F})  &&= \frac{1}{\pi^2}\int \int_{-\pi}^{\pi} \frac{-V_0  \cos\varphi  \cos\theta }{\sqrt{2\xi^{-2}+2k_F^2\left[1-\cos(\theta-\varphi)\right]} - V_0 \rho_{2D}}
d\theta d\varphi,
 \end{eqnarray} and then \begin{eqnarray} \label{eqD}{\lambda_{\textrm{eff}}^{RPA}}^{(1)}  = \frac{2\sqrt{2}}{\pi}\; V_0 \xi \; \mathcal{I}(X,Y),
\end{eqnarray} with $Y=V_0\rho_{2D}\xi/\sqrt{2}$, and \small \begin{eqnarray}\nonumber
\mathcal{I}(X,Y)&&=  \bigg\{\frac{ (1 + 2 X^2 - Y^2)^{3/2} \;K[\frac{2 X^2}{1 + 2 X^2}] +
      Y \bigg(\frac{\pi}{2} \sqrt{\frac{1 + 2 X^2}{1 - Y^2}} (1 + 2 X^2 - Y^2) +
          Y \sqrt{1 + 2 X^2 - Y^2} \;\Pi[\frac{2 X^2}{1 + 2 X^2 - Y^2}, \frac{2 X^2}{
           1 + 2 X^2}]\bigg)}{\sqrt{1 + 2 X^2} (1 + 2 X^2 - Y^2)^{3/2}}  \\ \nonumber  &&+\frac{\sqrt{(1 + 2 X^2) (1 + 2 X^2 - Y^2) (1 - Y^2)} \;E[\frac{2 X^2}{1 + 2 X^2}] -
       \sqrt{\frac{1 - Y^2}{1 + 2 X^2}} (1 + 2 X^2 - Y^2)^{3/2}
       \; K[\frac{2 X^2}{1 + 2 X^2}]}{X^2 \sqrt{(1 - Y^2) (1 + 2 X^2 - Y^2)}}
           \\ &&+\frac{
      Y \bigg(-\frac{\pi}{2} \left(1 + 2 X^2 - Y^2 -
            \sqrt{(1 - Y^2) (1 + 2 X^2 - Y^2)}\right) -
          Y \sqrt{\frac{(1 - Y^2) (1 + 2 X^2 - Y^2)}{1 + 2 X^2}}
           \; \Pi[\frac{2 X^2}{1 + 2 X^2 - Y^2},\frac{2 X^2}{
           1 + 2 X^2}]\bigg)}{X^2 \sqrt{(1 - Y^2) (1 + 2 X^2 - Y^2)}}
            \bigg\},
\end{eqnarray} \normalsize where $\Pi[X,Y]$ is the complete elliptic integral of the third kind.  One can estimate the RPA gap correction comparing the minima in Fig.~\ref{fig3}, which shows the
profile of $\mathcal{F}(X)$ and $\mathcal{I}(X,Y)$ in a broad range of $X$ ($Y \sim 0.05/X$, since we consider $n_B$ as the only tunable parameter).

\subsection{Phonon lifetime}

The phonon lifetime ($\tau$) due to particle-hole excitation is
\begin{eqnarray}
\frac{1}{\tau}=-2 \:\mathcal{I}m\Sigma(\mathbf{q},\nu),
\end{eqnarray} where \begin{eqnarray}
\Sigma(\mathbf{q},i \nu) = g_{FB}^2n_0 V_{\mathbf{q}}^2 \mathcal{D}_0(\mathbf{q},i\nu)^2
P(\mathbf{q},i \nu),
\end{eqnarray} \begin{figure}[!ht]
  \centering
  \includegraphics[width=0.3\textwidth]{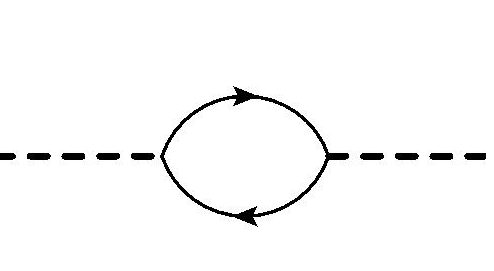}
  \caption{Polarization bubble in the phonon propagator.}
   \label{fig4}
\end{figure}  as shown in Fig.\ref{fig4}, \\ \\
with the polarization bubble \begin{eqnarray}
P(\mathbf{q},i \nu) =  \frac{1}{\beta S}\sum_{n,\mathbf{p}} \mathcal{G}_0(\mathbf{p},i\omega_n)\mathcal{G}_0(\mathbf{p}+\mathbf{q},i\omega_n+i\nu).
\end{eqnarray} Since we have  \begin{eqnarray}  \nonumber
\mathcal{I}m\: P(\mathbf{q},\nu)= -\frac{m_F^2}{\pi k_F^2}\left\{\Theta\left(1-\bigg|\frac{q}{2k_F}+\frac{m_F \nu}{q k_F}\bigg|\right)\sqrt{\frac{2 q^2 \mu}{m_F}-(\varepsilon_{\mathbf{q}}+\nu)^2}- \Theta\left(1-\bigg|\frac{q}{2k_F}-\frac{m_F \nu}{q k_F}\bigg|\right)\sqrt{\frac{2 q^2 \mu}{m_F}-(\varepsilon_{\mathbf{q}}-\nu)^2}\right\}, \\
\end{eqnarray} with the Fermi energy $\mu=k_F^2/2m_F$, then $\tau =\infty$ for $\nu=0$ (static limit considered in the paper). 

\subsection{Vertex correction}
We still have to deal with the $4th$-order vertex-correction in Fig.~\ref{fig4}(c) \begin{eqnarray} \nonumber
&&\Gamma_{V}^{(4)}(\{\mathbf{k}_i,\nu_i\})=
 -\frac{g_{FB}^4 n_B^2 }{V^2} \delta_{\mathbf{k}_1+\mathbf{k}_2,\mathbf{k}_3+\mathbf{k}_4}\beta \delta_{\nu_1+\nu_2,\nu_3+\nu_4} \prod_{i=1...4} \mathcal{G}_0(\mathbf{k}_i,\nu_i) \\ && \times
 \sum_{\mathbf{p},q_{2z},q_{4z}} V^2_{\mathbf{q}_2}V^2_{\mathbf{q}_4} \mathcal{D}_0(\mathbf{q}_2,i\nu_2-i\nu_3)\sum_n \mathcal{D}_0(\mathbf{q}_4,i\nu_3-i\nu_n)
\mathcal{G}_0(\mathbf{p}+\mathbf{k}_4-\mathbf{k}_1,\nu_4-\nu_1+\nu_n)\mathcal{G}_0(\mathbf{p},\nu_n), \end{eqnarray} with $\mathbf{q}_2 = (\mathbf{k}_2-\mathbf{k}_3,q_{2z})$ and $\mathbf{q}_4 =
(\mathbf{k}_3-\mathbf{p},q_{4z})$. \\

That leads to \begin{eqnarray} \nonumber \label{vertx}
&& \Gamma_{V}^{(4)}(\{\mathbf{k}_i,\nu_i\}) =   \frac{2 g_{FB}^4 n_B^2 m_B}{\sqrt{|\mathbf{k}_1-\mathbf{k}_4|^2+2\xi^{-2}}} \frac{1}{V}\frac{\beta}{S}
\delta_{\mathbf{k}_1+\mathbf{k}_2,\mathbf{k}_3+\mathbf{k}_4} \delta_{\nu_1+\nu_2,\nu_3+\nu_4} \prod_{i=1...4} \mathcal{G}_0(\mathbf{k}_i,\nu_i)
\\ && \times \sum_{\mathbf{p},q_z} \frac{q}{\sqrt{q^2+2\xi^{-2}}} \bigg[\frac{1}{(\omega_{\mathbf{q}}+\epsilon_{\mathbf{p}})(\omega_{\mathbf{q}}+\epsilon_{\mathbf{p}+\mathbf{k}_4-\mathbf{k}_1})}
+\frac{4n_F(\epsilon_{\mathbf{p}})\omega_{\mathbf{q}}}{(\epsilon_{\mathbf{p}}-\epsilon_{\mathbf{p}+\mathbf{k}_4-\mathbf{k}_1})(\omega^2_{\mathbf{q}}-\epsilon^2_{\mathbf{p}})}  \bigg],  \end{eqnarray}
with $\omega_{\mathbf{q}}=\frac{q}{2m_B}\sqrt{q^2+2\xi^{-2}}$ and $\mathbf{q}\equiv(\mathbf{k}_3-\mathbf{p},q_z)$.
\begin{figure}[!ht]
  \centering
  \includegraphics[width=16.5cm]{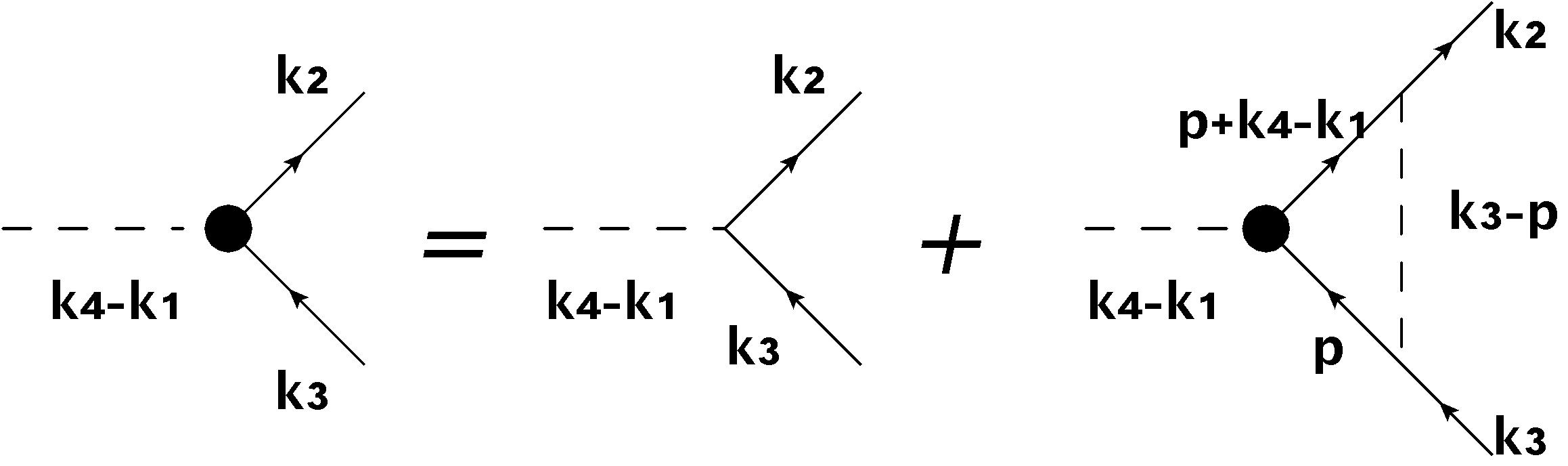}
  \caption{Feynman diagram for the self-consistent vertex equation in Eq.~(\ref{vertexEQ}). \label{fig5} }
\end{figure}
\subsection{Self-consistent vertex equation}
Summing the ladder series as shown in Fig.~\ref{fig5}, we derive the self-consistent vertex equation  \begin{eqnarray}  \label{vertexEQ}\nonumber
&&\lambda(\mathbf{k}_3,\mathbf{k}_4-\mathbf{k}_1;i\nu_3,i\nu_4-i\nu_1)=\lambda_0(\mathbf{k}_3,\mathbf{k}_4-\mathbf{k}_1;i\nu_3,i\nu_4-i\nu_1) -g_{FB}^2n_B\frac{1}{V \beta}\sum_{\mathbf{p},q_z}\sum_{n}
V^2_{\mathbf{q}} \mathcal{D}_0(\mathbf{q},i\nu_3-i\nu_n)  \\ \nonumber &&\times \mathcal{G}_0(\mathbf{p},i\nu_n)\mathcal{G}_0(\mathbf{p}+\mathbf{k}_4-\mathbf{k}_1,i\nu_n+i\nu_4-i\nu_1)
\lambda(\mathbf{p},\mathbf{k}_4-\mathbf{k}_1;i\nu_n,i\nu_4-i\nu_1). \\
\end{eqnarray} After considering $\epsilon_{\mathbf{p}}=\varepsilon_{\mathbf{p}}-\mu \sim 0$ and $\lambda=\lambda(|\mathbf{k}_4-\mathbf{k}_1|)$, again for zero external frequencies $\nu_i=0$, we can deal
with the remaining sum    \begin{eqnarray} \nonumber
&& \tilde{\Pi}(\mathbf{k}_4,\mathbf{k}_1) =  \frac{1}{V\beta} \sum_{\mathbf{p},q_z}\sum_{n} V^2_{\mathbf{q}} \mathcal{D}_0(\mathbf{q},-i\nu_n)
\mathcal{G}_0(\mathbf{p},i\nu_n)\mathcal{G}_0(\mathbf{p}+\mathbf{k}_4-\mathbf{k}_1,i\nu_n)  = \\ \nonumber
 &&  = \frac{1}{V} \sum_{\mathbf{p},q_z} \frac{q}{\sqrt{q^2+2\xi^{-2}}}\;
 \frac{4n_F(\epsilon_{\mathbf{p}})}{\varepsilon_{\mathbf{p}}-\varepsilon_{\mathbf{p}+\mathbf{k}_4-\mathbf{k}_1}}\frac{1}{\omega_{\mathbf{q}}} \\ \nonumber
 && =  {-16 m_B m_F}\int \frac{d^2p}{(2\pi)^2} \int \frac{dq_z}{2\pi} \frac{1}{k_F^2+p^2-2 k_F p \cos(\theta-\theta_3)+q_z^2+2\xi^{-2} } \; \frac{1}{|\mathbf{k}_4-\mathbf{k}_1|^2+2 k_F p
 [\cos(\theta-\theta_4)-\cos(\theta-\theta_1)]}  \\  \nonumber
 && =  -\frac{2 m_B m_F}{\pi^2}\int_0^{k_F} p dp \int_0^{2\pi} d\theta \frac{1}{\sqrt{k_F^2+p^2-2 k_F p \cos(\theta-\theta_3)+2\xi^{-2} }} \; \frac{1}{|\mathbf{k}_4-\mathbf{k}_1|^2+2 k_F p
 [\cos(\theta-\theta_4)-\cos(\theta-\theta_1)],} \\ \end{eqnarray}  with the additional external momenta constraint $\theta_2-\theta_1=\pi$ and $\theta_4-\theta_3=\pi$, and $|\mathbf{k}_4-\mathbf{k}_1|^2 = 2 k_F^2 [1 - \cos(\theta_1 - \theta_4)]$.

We finally obtain the vertex correction after substituting the angular momentum expansion \begin{eqnarray}
\lambda(|\mathbf{k}_4-\mathbf{k}_1|) = \sum_{\ell}  \lambda^{(\ell)}(k_F) \cos[\ell (\theta_4 -\theta_1)]
\end{eqnarray} in Eq.~(\ref{vertexEQ}), to obtain the decoupled equation for the projection $\ell=1$ \begin{eqnarray}  \label{vertexEQl}
 \lambda^{(1)}(k_F) =\lambda_0^{(1)}(k_F) -g_{FB}^2n_B \; \lambda^{(1)}(k_F) \; \Pi^{(1)}(k_F),
\end{eqnarray} where $\lambda_0^{(1)} =  V_{\textrm{eff}}^{(1)}$, as calculated in the main text, and \begin{eqnarray}  \label{Pi}
\Pi^{(1)}(k_F) &&= \frac{1}{\pi^2} \int_{-\pi}^{\pi} d\theta_1 \;{\cos\theta_1} \int_{-\pi}^{\pi} d\theta_4 \; {\cos\theta_4}\; \tilde{\Pi}(\mathbf{k}_4,\mathbf{k}_1) \cos(\theta_4-\theta_1).
\end{eqnarray}

After considering $p=k_F$ in the integrant of Eq.~(\ref{Pi}), we have to deal with the angular integrals \begin{eqnarray} \nonumber
\Pi^{(1)}(k_F)&&=  - \frac{m_B m_F}{2\sqrt{2}\pi^4k_F}\int_{-\pi}^{\pi} d\theta_1 \;{\cos\theta_1} \int_{-\pi}^{\pi} d\theta_4 \; {\cos\theta_4} \int_0^{2\pi} d\theta \frac{ \cos(\theta_4-\theta_1)}{\sqrt{1+
\cos(\theta-\theta_4)+(\xi k_F)^{-2}}} \\ \nonumber &&\times \frac{1}{1 - \cos(\theta_1 - \theta_4)+\cos(\theta-\theta_4)-\cos(\theta-\theta_1)} \\
&&\sim  \frac{1}{\sqrt{2}\pi^2}\frac{m_B m_F}{k_F^2\xi}\frac{\mathcal{J}[k_F\xi]}{\sqrt{1+2k_F^2\xi^2}},
\end{eqnarray} with \begin{eqnarray} \mathcal{J}[X]=(1+2X^2) E\left[1 - \frac{1}{1+2X^2}\right]-(1+X^2) K\left[1 - \frac{1}{1+2X^2}\right]. \end{eqnarray} \\ Then, from Eq.~(\ref{vertexEQl}) we finally
get \begin{eqnarray}  \label{vertex2}
 \lambda^{(1)}(k_F) = \frac{\frac{4\sqrt{2}}{\pi}\; g_{FB}^2 n_B m_B  \xi \; \mathcal{F}(k_F\xi)}{1+ g_{FB}^2 n_B \frac{1}{\sqrt{2}\pi^2}\frac{ m_B
 m_F}{k_F^2\xi}\frac{\mathcal{J}[k_F\xi]}{\sqrt{1+2k_F^2\xi^2}}}.
\end{eqnarray}

\end{document}